\documentclass[aps,preprint,superscriptaddress,groupedaddress,nofootinbib]{revtex4}

\usepackage{graphicx}
\usepackage[normalem]{ulem}
\usepackage[export]{adjustbox}
\usepackage{float}
\usepackage{amsmath}
\usepackage{amssymb}
\usepackage[rightcaption]{sidecap}
\usepackage{hyperref}
\usepackage{xcolor}
\graphicspath{ {./images/} }

\def\figureautorefname~#1\null{Fig.\,#1\null}
\def\tableautorefname~#1\null{Tab.\,#1\null}

\def\equationautorefname~#1\null{Eq.\,(#1)\null}

\allowdisplaybreaks[4]

\begin{document}

\title{Electroweak phase transition in the 2HDM: \\Collider and gravitational wave complementarity}
\author{Dorival Gon\c{c}alves}
\email{dorival@okstate.edu}
\affiliation{Department of Physics, Oklahoma State University, Stillwater, OK, 74078, USA}
\author{Ajay Kaladharan}
\email{kaajay@ostatemail.okstate.edu}
\affiliation{Department of Physics, Oklahoma State University, Stillwater, OK, 74078, USA}
\author{Yongcheng Wu}
\email{ywu@okstate.edu}
\affiliation{Department of Physics, Oklahoma State University, Stillwater, OK, 74078, USA}

\begin{abstract}
The knowledge of the Higgs potential is  crucial for understanding the origin of mass and the thermal history of our Universe. We show how collider measurements and observations of stochastic gravitational wave signals can complement each other to explore the rich scalar potential in the two Higgs doublet model (2HDM). Accounting for theoretical and current experimental constraints, we analyze the key ingredients in the shape of the Higgs potential driving the transition from a smooth crossover to a strong first-order phase transition ($\xi_c>1$), focusing on the barrier formation and the upliftment of the true vacuum. In particular, we observe that the $\xi_c>1$ regime is favored for lower scalar masses, providing strong additional motivation for collider searches. We contrast the dominant collider signals at the HL-LHC (high-luminosity LHC) with observable gravitational wave signals at LISA. We obtain that the HL-LHC will be able to cover a vast range of the $\xi_c>1$ parameter space, with scalar decays to heavy fermions $(H,A,H^\pm\to tt, tb)$ being the most promising smoking gun signature of a strong first-order electroweak phase transition in the 2HDM. In contrast, LISA provides sensitivity to only a limited region of the parameter space.
\end{abstract}

\maketitle
\flushbottom
\tableofcontents
\newpage

\section{Introduction}

The structure of the Higgs potential is deeply connected with the origin of mass and thermal history of our Universe. When the Universe was cooling down, at a temperature of order 100~GeV, it went through a transition from a symmetric phase to an electroweak (EW) broken phase, where the Higgs field(s) acquired nonvanishing vacuum expectation values. The evolution of this process strongly depends on the shape of the Higgs potential. Distinct profiles for the Higgs potential result in contrasting courses for the electroweak symmetry breaking (EWSB) in the early Universe, ranging from the smooth crossover transition in the Standard Model (SM), with the observed 125~GeV Higgs boson~\cite{Kajantie:1996mn}, to the strong first-order phase transition, with new physics.

The dynamics of the electroweak phase transition (EWPT) could have profound consequences for particle physics and cosmology. Most notably, it may be behind the matter and antimatter asymmetry puzzle. This asymmetry can be quantitatively featured by the baryon-to-photon ratio measurement $n_B/n_\gamma\approx 6\times 10^{-10}$~\cite{Planck:2015fie}, which is several orders of magnitude larger than the expected for the symmetric scenario, indicating an asymmetry in the early Universe between baryons and antibaryons. The bulk of the antibaryons have been annihilated in the thermal history, resulting in the large present density of baryons. The ingredients required to generate the baryon asymmetry of the Universe are theoretically well understood and summarized by the three Sakharov conditions~\cite{Sakharov:1967dj}. They impose that our fundamental theory displays baryon number violation, $C$ along with $CP$ violation, and departure from thermal equilibrium. Whereas the SM satisfies baryon number and $C$ violation, the source of $CP$ violation from the Cabibbo-Kobayashi-Maskawa (CKM) matrix is too small, and the observed Higgs mass $m_h=125$~GeV is too high to generate out-of-equilibrium conditions from a strong first-order phase transition~\cite{Huet:1994jb,Kajantie:1996mn}. Thus, baryogenesis requires physics beyond the SM to grant these missing ingredients. In the present work, we focus on the latter problem, generating the out-of-equilibrium conditions at the electroweak scale~\cite{Trodden:1998ym,Cohen:1993nk,Carena:1996wj,Morrissey:2012db,Goncalves:2023svb}.

The transmutation of the EWPT pattern from smooth crossover to strong first-order phase transition usually requires new degrees of freedom around the EW scale, with sizable interactions with the Higgs boson~\cite{Ramsey-Musolf:2019lsf}. Therefore, it generally accommodates beyond-the-SM scenarios with exciting phenomenological prospects, both at collider and gravitational wave (GW) experiments.
At colliders, the Higgs pair production $pp\to hh$ usually plays a leading role in this discussion, as it grants a direct probe of the Higgs potential~\cite{Eboli:1987dy,Plehn:1996wb,No:2013wsa}. It provides access to the triple Higgs coupling via nonresonant Higgs pair production, as well as to the interactions of the SM Higgs boson with new heavy scalars through resonant di-Higgs searches. The high-luminosity LHC projections indicate that the triple Higgs coupling will be bounded to $0.1<\lambda_{h^3}/\lambda_{h^3}^{\rm SM}<2.3$ at 95\% confidence level (C.L.)~\cite{Cepeda:2019klc}. Resonant searches are also a main focus, leading to significant limits~\cite{ATLAS:2018rnh}.
Complementing the collider searches, the space-based GW experiments, such as LISA~\cite{Audley:2017drz}, will provide a new window to the Higgs potential. First-order phase transitions that emerge from a scalar field tunneling through an energy barrier in the potential generate a significant source of gravitational waves. The corresponding signal spectrum displays a characteristic peak associated with the temperature at which the phase transition occurred. For phase transitions at the EW scale, this leads to a GW spectrum around the mHz frequency, after redshifting the signal to the present time~\cite{Grojean:2006bp,Caprini:2019egz}. This prompts exciting prospects to access the nature of EWPT at LISA, as this is precisely the frequency band to which this experiment is sensitive.

In this work, we study the EWPT pattern in the two Higgs doublet model (2HDM)~\cite{Dorsch:2013wja,Basler:2016obg,Dorsch:2016tab,Bernon:2017jgv,Dorsch:2017nza,Andersen:2017ika,Kainulainen:2019kyp,Su:2020pjw,Davoudiasl:2021syn,Biekotter:2021ysx,Aoki:2021oez}, where the SM is augmented by an extra doublet. Instead of focusing on benchmarks or a particular parameter space region, a general scan is performed on the theoretically and experimentally allowed parameter space. We divide our analysis into two main stages. First, we scrutinize the new physics modifications to the shape of the Higgs potential that lead to a strong first-order phase transition. We devote particular attention to the barrier formation and to the true vacuum upliftment with respect to the SM case.
The obtained results work as a guide for the phenomenological studies derived in the second part of the manuscript, where we perform the respective analysis at both the HL-LHC and LISA. In this last part, besides the commonly discussed channel $A\to ZH$, other promising channels that can cover the EWPT parameter space are also investigated. These studies highlight the leading collider signatures for first-order EWPT at the LHC, as well as the complementarity of probes between collider and GW experiments.

The structure of this paper is as follows. In~\autoref{sec:2HDM}, we briefly describe the 2HDM. The one-loop effective potential at finite temperature is discussed in~\autoref{sec:Veff}. It is followed by an introduction of EWPT and GW signals in~\autoref{sec:EWPT_GW}. In~\autoref{sec:Vshape}, we discuss how the shape of the potential will affect the EWPT, focusing on the barrier formation and vacuum upliftment. In~\autoref{sec:Collider_GW}, inspired by our shape analysis, we tailor the collider studies to the most promising channels. In addition, we derive the sensitivity to the corresponding GW signals generated in the early Universe. Finally, we summarize in~\autoref{sec:summary}. Some useful relations for the parameters in 2HDM are listed in Appendix~\ref{app:parameters}.

\section{Two Higgs Doublet Model}
\label{sec:2HDM}

The two Higgs doublet model displays one of the most minimalistic extensions of the SM that is compatible with the current experimental constraints~\cite{Branco:2011iw}. In this work, we consider the $CP$-conserving 2HDM with a softly broken $\mathbb{Z}_2$ symmetry.\footnote{{Baryogenesis requires physics beyond the SM to generate new sources of CP violation and out-of-equilibrium conditions. In the present work, we focus on the latter issue.}} The tree-level potential is given by
\begin{align}
\begin{split}
    V(\Phi_1,\Phi_2) =& m_{11}^2\Phi_1^\dagger\Phi_1 + m_{22}^2\Phi_2^\dagger\Phi_2 - m_{12}^2(\Phi_1^\dagger\Phi_2 + {\rm H.c.}) + \frac{\lambda_1}{2}(\Phi_1^\dagger\Phi_1)^2 + \frac{\lambda_2}{2}(\Phi_2^\dagger\Phi_2)^2 \\
            & + \lambda_3(\Phi_1^\dagger\Phi_1)(\Phi_2^\dagger\Phi_2) + \lambda_4(\Phi_1^\dagger\Phi_2)(\Phi_2^\dagger\Phi_1) + \frac{\lambda_5}{2}\left((\Phi_1^\dagger\Phi_2)^2 + {\rm H.c.}\right),
\label{equ:v_tree}
\end{split}
\end{align}
where the mass terms $m_{11}^2,~m_{22}^2,$ and $m_{12}^2$ along with the couplings $\lambda_1...\lambda_5$ are real parameters from Hermiticity and $CP$-conservation. The $\mathbb{Z}_2$ symmetry, $\Phi_1\to\Phi_1$ and $\Phi_2\to-\Phi_2$, softly broken by $m_{12}^2$, guarantees the absence of dangerous tree-level flavor-changing neutral currents(FCNCs)~\cite{PhysRevD.15.1958,PhysRevD.15.1966}. After EWSB, the neutral components of the two  $SU(2)_L$ doublets develop vacuum expectation values (VEVs). Expanding around the  VEVs $\tilde{\omega}_i$, the scalar doublets $\Phi_i$ may be written as
\begin{align}
    \Phi_1 = \left(\begin{array}{c}
        \phi_1^+\\
        \frac{\tilde{\omega}_1 + \phi_1^0 + i \eta_1}{\sqrt{2}}
    \end{array}\right)
    \hspace{0.5cm} \text{and} \hspace{0.5cm}
        \Phi_2 = \left(\begin{array}{c}
        \phi_2^+\\
        \frac{\tilde{\omega}_2 + \phi_2^0 + i \eta_2}{\sqrt{2}}
    \end{array}\right)\,,
\end{align}
where the zero-temperature vacuum expectation values $v_i\equiv \tilde{\omega}_i|_{T=0}$ are connected to the SM VEV by $v_1^2+v_2^2=v^2\approx (246\,{\rm GeV})^2$.

The $CP$-conserving 2HDM leads to five physical mass eigenstates in the scalar sector: two $CP$-even neutral scalars $h$ and $H$, a neutral $CP$-odd scalar $A$, and a charged scalar pair $H^\pm$. The relation between the mass and gauge eigenstates is established by the rotation angle $\beta$ for the charged and $CP$-odd sectors, where $\tan\beta\equiv v_2/v_1$, and by the mixing angle $\alpha$ in the $CP$-even sector
\begin{align}
    \left(\begin{array}{c}
        G^\pm\\
        H^\pm
    \end{array}\right) =
    \mathcal{R}(\beta)
    \left(\begin{array}{c}
        \phi_1^\pm\\
        \phi_2^\pm
    \end{array}\right),\,\,\,
    \left(\begin{array}{c}
        G^0\\
        A
    \end{array}\right)= \mathcal{R}(\beta)
    \left(\begin{array}{c}
        \eta_1\\
        \eta_2
    \end{array}\right),\,\,\,
    \left(\begin{array}{c}
        H\\
        h
    \end{array}\right)= \mathcal{R}(\alpha)
    \left(\begin{array}{c}
        \phi_1^0\\
        \phi_2^0
    \end{array}\right).
\end{align}
The rotation matrix is defined as
\begin{align}
     \mathcal{R}(x)=\left(\begin{array}{cc}
        c_x  & s_x \\
        -s_x & c_x
    \end{array}\right)\,,
\end{align}
with $s_{x}\equiv \sin x$ and $c_{x}\equiv \cos x$. $G^\pm$ and $G^0$ represent the charged and neutral massless Goldstone bosons.

Instead of the eight parameters in the Higgs potential $m_{11}^2$, $m_{22}^2$, $m_{12}^2$, $\lambda_1...\lambda_5$, a more convenient choice of parameters is
\begin{equation}
  \tan\beta,~\cos (\beta-\alpha),~m_{12}^2,~v,~m_h,~m_H,~m_A,~m_{H^\pm} \,.
\end{equation}
The conversion between these two sets of parameters can be found in Appendix~\ref{app:parameters}. The parameters $t_\beta\equiv\tan\beta$ and $c_{\beta-\alpha}\equiv\cos(\beta-\alpha)$ are of critical phenomenological importance. They control the coupling strength of scalar particles to fermions and gauge bosons. Given the current experimental constraints, a particular relevant regime is the alignment limit $c_{\beta-\alpha}=0$~\cite{Gunion:2002zf}, where the 125~GeV $CP$-even scalar Higgs boson couples to SM particles precisely as the SM Higgs boson.

In general, there are four types of $\mathbb{Z}_2$ charge assignments in the Yukawa sector that avoid FCNC at tree level. In this work, we focus on the type-I and type-II scenarios. In the first case, all fermions couple only to $\Phi_2$, whereas in the latter, only the up quarks couple with  $\Phi_2$, leaving the down quarks and charged leptons to couple with $\Phi_1$. For both types I and II, we perform a uniform scan over the parameter space region,
\begin{align}
    \tan\beta &\in (0.8,25)\,,                &m_{12}^2&\in(10^{-3},10^5)\,{\rm GeV}^2\,,  &m_H&\in(150,1500)\rm\,GeV \,,
    \nonumber \\
    \cos(\beta-\alpha)&\in(-0.3,0.3)\,,  & m_A&\in(150,1500)\,{\rm GeV}\,,                     &m_{H^\pm}&\in(150,1500)\,{\rm GeV}.
     \label{eq:param_scan}
\end{align}
The observed 125~GeV Higgs boson is identified with the $h$ scalar. The parameter space scan is performed with {\tt ScannerS}~\cite{Coimbra:2013qq,Muhlleitner:2020wwk}. Using this framework, we impose the constraints from perturbative unitarity~\cite{Lee:1977eg,Kanemura:1993hm,Ginzburg:2005dt}, boundedness from below~\cite{Ivanov:2018jmz}, vacuum stability~\cite{Hollik:2018wrr,Ferreira:2019iqb}, electroweak precision, and flavor constraints. In addition, {\tt HiggsBounds} and {\tt HiggsSignals} are used to incorporate the searches for additional scalars as well as the constraints from the 125~GeV Higgs boson measurements~\cite{Bechtle:2020pkv,Bechtle:2020uwn}.

\section{One-Loop Effective Potential at Finite Temperature}
\label{sec:Veff}
To study the electroweak phase transition in the early Universe, we use the loop-corrected effective potential at finite temperature. In addition to the tree-level potential $V_0$ from~\autoref{equ:v_tree}, the effective potential displays one-loop corrections at zero temperature from the Coleman-Weinberg potential $V_{CW}$ and counterterms $V_{\rm CT}$. Finite-temperature corrections $V_T$ are also included. The effective potential reads
\begin{align}
    V_{\rm eff} = V_0 + V_{CW} + V_{CT}+ V_T\,.
    \label{eq:Veff}
\end{align}

The Coleman-Weinberg potential can be written in the Landau gauge as~\cite{PhysRevD.7.1888}
\begin{align}
    V_{CW} &= \sum_i \frac{n_i}{64\pi^2}m_i^4(\Phi_1,\Phi_2)\left[\log \left(\frac{m_i^2(\Phi_1,\Phi_2)}{\mu^2}\right)-c_i\right]\,,
    \label{eq:CW}
\end{align}
where the index $i$ sums over all particles in the thermal bath with field-dependent mass $m_i(\Phi_1,\Phi_2)$, namely, massive gauge bosons, longitudinal photon, Higgs bosons, Goldstone bosons, and fermions. $n_i$ denotes the number of degrees of freedom for particle $i$, with $n_i>0$ for bosons and $n_i<0$ for fermions. The various constants $c_i$ depend on the renormalization scheme adopted. Following the $\overline{\text{MS}}$ scheme, we set $c_i$ to 5/6 for gauge bosons and 3/2 otherwise. Finally, the renormalization scale $\mu$ is fixed at the zero-temperature VEV, $\mu=v(T=0)\approx 246~$GeV.\footnote{
A renormalization group improved calculation can be taken into account for a further refined estimation~\cite{Chiang:2017nmu}. For the scale-dependence problem at finite temperature, we refer to Ref.~\cite{Gould:2021oba} for a detailed discussion.}

In general, the one-loop Coleman-Weinberg corrections shift the scalar masses and mixing angles with respect to the tree-level potential. To optimize our parameter scan, we adopt a renormalization prescription that enforces these parameters to match with their tree-level values~\cite{Camargo-Molina:2016moz,Basler:2016obg}. In this setup, the counterterm part of the potential can be written as
\begin{align}
	V_{CT} =& \delta m_{11}^2\Phi_1^\dagger\Phi_1 + \delta m_{22}^2\Phi_2^\dagger\Phi_2 - \delta m_{12}^2(\Phi_1^\dagger\Phi_2 + {\rm H.c.}) + \frac{\delta \lambda_1}{2}(\Phi_1^\dagger\Phi_1)^2 + \frac{\delta\lambda_2}{2}(\Phi_2^\dagger\Phi_2)^2 \nonumber \\
            & + \delta\lambda_3(\Phi_1^\dagger\Phi_1)(\Phi_2^\dagger\Phi_2) + \delta\lambda_4(\Phi_1^\dagger\Phi_2)(\Phi_2^\dagger\Phi_1) + \frac{\delta\lambda_5}{2}\left((\Phi_1^\dagger\Phi_2)^2 + {\rm H.c.}\right)\,,
\label{eq:CT}
\end{align}
with the following on-shell renormalization conditions at zero temperature:
\begin{align}
\partial_{\phi_i}(V_{CW}+V_{CT})|_{\omega=\omega_\text{tree}}=0 \,,
\label{eq:ren1}\\
\partial_{\phi_i}\partial_{\phi_j}(V_{CW}+V_{CT})|_{\omega=\omega_\text{tree}}=0\,.
\label{eq:ren2}
\end{align}
The fields $\phi_i$ ($i=1,...,8$) denote  scalar components from the $\Phi_1$ and $\Phi_2$ doublets, $\omega$ generically represents the $\omega_i$ values, and $\omega_\text{tree}$ generically stands for the minimum of the tree-level potential for the $\phi_i$ fields.
We follow the prescription of Ref.~\cite{Camargo-Molina:2016moz} to consistently calculate the first and second derivatives of $V_{CW}$.
The first renormalization condition~\autoref{eq:ren1} imposes that the zero-temperature minimum is not shifted with respect to the tree-level value. Similarly, the second condition~\autoref{eq:ren2} ensures that the zero-temperature masses and mixing angles remain the same as their tree-level assignment.

The last term in~\autoref{eq:Veff}, the one-loop thermal corrections $V_T$, can be expressed as~\cite{Arnold:1992rz}
\begin{align}
    V_T&=\frac{T^4}{2\pi^2} \left[
    \sum_f n_f J_+\left(\frac{m_f^2}{T^2}\right)+
    \sum_{\mathcal{V}_T} n_{\mathcal{V}_T} J_-\left(\frac{m_{\mathcal{V}_T}^2}{T^2}\right)
    +\sum_{\mathcal{V}_L} n_{\mathcal{V}_L}  J_-\left(\frac{m_{\mathcal{V}_L}^2}{T^2}\right) \right] \nonumber\\
    &-\frac{T^4}{2\pi^2}\sum_{\mathcal{V}_L} \frac{\pi}{6} \left(\frac{\overline{m}^3_{\mathcal{V}_L}}{T^3}-\frac{m_{\mathcal{V}_L}^3}{T^3}\right) \,,
    \label{eq:VT}
\end{align}
where the sum extends over fermions $f$ and bosons, with the latter subdivided in {transverse modes of gauge boson}  $\mathcal{V}_T=W_T,Z_T$ and {longitudinal modes of gauge bosons and scalars} $\mathcal{V}_L=W_L,Z_L,\gamma_L,\Phi^0,\Phi^{\pm}$. The resummation of the $n=0$ Matsubara modes of $\mathcal{V}_L$ results in thermal corrections to their masses~\cite{Matsubara:1955ws,Quiros:1999jp}. The second line in~\autoref{eq:VT} indicates the Daisy contributions, where $\overline{m}_{\mathcal{V}_L}$ is the thermal Debye mass following the Arnold-Espinosa scheme~\cite{Arnold:1992rz,Basler:2016obg}. Lastly, the thermal functions for fermions $(J_+)$ and for bosons $(J_-)$ read
\begin{align}
    J_{\pm}(x)=\mp\int_0^\infty dy~ y^2 \log\left(1\pm e^{-\sqrt{y^2+x^2}} \right)\,.
\end{align}

{While the effective potential displays theoretical uncertainties arising from the choice of gauge parameter
~\cite{Patel:2011th,Wainwright:2011qy,Metaxas:1995ab,Garny:2012cg,Chiang:2017nmu,Arunasalam:2021zrs},
gauge-independent probes can be constructed  exploiting the Nielsen identities~\cite{Nielsen:1975fs}. These identities state that the gauge dependence vanishes at the extrema of the potential
\begin{equation}
    \frac{\partial V_\text{eff}(\Phi _1,\Phi _2,\xi )}{\partial \xi }=-C_i(\Phi _1,\Phi _2,\xi)\frac{\partial V_\text{eff}(\Phi _1,\Phi _2,\xi )}{\partial \phi_i }\,,
\end{equation}
where $\xi$ is the gauge fixing parameter. This motivates us to adopt two distinct methods in our manuscript for the phenomenological analyses. The first approach encompasses the calculation of the finite-temperature effective potential and the subsequent numerical scan. The second one focuses on the calculation of the vacuum upliftment at $T=0$. As we will highlight in Sec.~\ref{sec:vauum-uplift}, the upliftment of the true vacuum with respect to the symmetric one at zero temperature works as an effective probe of the strength of phase transition. Whereas the first method displays uncertainties rooted in the choice of  gauge parameter, the latter approach is gauge invariant, as guaranteed by the Nielsen identities~\cite{Dorsch:2017nza,Patel:2011th}.  Notice that we introduce extra counterterms to preserve the position of the EW vacuum, as well as the masses, at one-loop order. The phenomenological agreement between our numerical scan with the  profile derived from the vacuum upliftment will, in particular, show that the numerical scan is  well grounded, despite its uncertainties.}\footnote{Note that when the coupling in the scalar potential is large, as usually required by a strong first-order phase transition, one should check the reliability of the perturbative calculations. Lattice simulations for 2HDM are performed for two particular benchmark points in Ref.~\cite{Kainulainen:2019kyp}, where the authors made comparisons among different methods.  It is shown that the perturbative estimation of the strength of the phase transition $\xi_c\equiv v_c/T_c$ is close to the lattice results.}

\section{Electroweak Phase Transition and Gravitational Waves}
\label{sec:EWPT_GW}
The effective potential at finite temperature determines the dynamics of the phase transition. The two Higgs doublet model exhibits multiple phase transition processes. For successful baryogenesis, the sphaleron process inside the bubble should be heavily suppressed to prevent the net baryon number generated around the bubble wall from significant washout.
{This condition requires that the EWPT be of strong first order~\cite{Quiros:1999jp}}
\begin{equation}
   \xi_c\equiv \frac{v_c}{T_c} \gtrsim 1\,,
\end{equation}
where $v_c\equiv \sqrt{\omega_1^2+\omega_2^2} |_{T_c}$ is the Higgs VEV at the critical temperature $T_c$, which is defined when the would-be true vacuum and false vacuum are degenerate.
{The approximate inequality denotes the  theoretical uncertainty  in this condition~\cite{Patel:2011th}.}

The transition from false to true vacuum takes place via thermal tunneling. It results in the formation of bubbles of the broken phase that expand in the surrounding region of symmetric phase, converting the false vacuum into true vacuum. The tunneling probability can be written as~\cite{Linde:1980tt,Coleman:1977py}
\begin{equation}
    \Gamma (T)\approx T^4\left (\frac {S_3}{2\pi T}  \right )^{3/2}e^{-\frac {S_3}{T}}\,,
\end{equation}
where $S_3$ is the three-dimensional Euclidean action corresponding to the critical bubble
\begin{equation}
    S_3=4\pi\int_{0}^{\infty}{dr r^2\left [ \frac 12\left ( \frac {d\phi(r)}{dr} \right )^2+V(\phi,T) \right ]}\,.
\end{equation}
Here, the scalar field $\phi$ is the bubble profile of the critical bubble. It is obtained as a solution to the following differential equation
\begin{equation}
    \frac {d^2\phi}{dr^2}+\frac 2r \frac{d\phi}{dr}=\frac {dV(\phi,T)}{d\phi}\,, \quad   \text{with} \quad \lim_{r\rightarrow \infty}\phi(r)=0
    \quad \text{and} \quad \lim_{r\rightarrow 0}\frac {d\phi(r)}{dr}=0.
    \label{eq:tunneling}
\end{equation}
We use the publicly available code {\tt CosmoTransitions} to solve the differential equation and compute the Euclidean action $S_3$~\cite{Wainwright:2011kj}.

The nucleation temperature $T_n$ is defined by requiring approximately one bubble to nucleate per Hubble volume~\cite{Moreno:1998bq}
\begin{equation}
    \int_{T_n}^{\infty}\frac {dT}{T}\frac {\Gamma (T)}{H(T)^4}=1\,.
\end{equation}
For the electroweak phase transition, the above condition can be roughly approximated as~\cite{Quiros:1999jp}\footnote{Nucleation does not by itself guarantee percolation and completion of the phase transition~\cite{Goncalves:2024vkj}. For the GW sensitive points obtained in our analysis, however, we find $\alpha=\mathcal{O}(0.1\text{--}1)$ and $\beta/H_n=\mathcal{O}(10\text{--}10^3)$, indicating that the relevant parameter space is not characterized by extreme supercooling or parametrically slow transitions. We therefore adopt the approximation $T_p\simeq T_n$ for the GW analysis~\cite{Goncalves:2023svb}.}
\begin{align}
\frac {S_3(T)}{T}\approx 140\,.
\end{align}

One of the important consequences of the strong first-order EWPT is the production of stochastic gravitational waves.
The GW signals from phase transition have three main sources:  collision of the vacuum bubbles, sound waves, and turbulence in the plasma. For each source, the GW spectrum can be expressed as numerical functions in terms of two important parameters determined from the phase transition dynamics~\cite{Grojean:2006bp,Caprini:2015zlo}. The first parameter is $\alpha\equiv \epsilon/\rho_{rad}$, the latent heat released in the phase transition $(\epsilon)$  to the radiation energy density $(\rho_{rad})$. The latent heat $\epsilon$ and $\rho_{rad}$ are obtained from
\begin{align}
 \epsilon = \Delta\left(- V_{\rm eff} + T\frac{\partial V_{\rm eff}}{\partial T}\right)_{T=T_n} \quad \text{and}\quad \rho_{\rm rad} = \frac{\pi^2}{30}g_\star T_n^4\,,
\end{align}
where $\Delta$ means the difference between the stable and metastable minima, and
$g_\star$ is the number of relativistic degrees of freedom in the plasma.
The second key parameter characterizing the spectrum of gravitational waves is the inverse time duration of the phase transition $\beta/H_n$.
This quantity is defined as
\begin{align}
	\frac{\beta}{H_n} &\equiv T_n\frac{d}{dT}\left.\left(\frac{S_3}{T}\right)\right|_{T=T_n}\,,
\end{align}
where $H_n$ denotes the Hubble constant at the nucleation temperature $T_n$. Strong GW signals are typically associated with large latent heat release (large $\alpha$) and slow phase transition (small $\beta/H_n$).

Finally, to estimate the sensitivity of GW experiments, we adopt the signal-to-noise ratio (SNR) measure~\cite{Caprini:2015zlo}
\begin{align}
\mathrm{SNR}=\sqrt{\mathcal{T} \int_{f_{\min }}^{f_{\max }} d f\left[\frac{h^{2} \Omega_{\mathrm{GW}}(f)}{h^{2} \Omega_{\mathrm{Sens}}(f)}\right]^{2}}\,,
\end{align}
where $\Omega_{\rm Sens}$ is the sensitive curve of the considered GW experiment~\cite{Audley:2017drz} and $\mathcal{T}$ is associated with the duration of the mission.  In the present study, we  focus on the LISA experiment as a benchmark, assuming $\mathcal{T}=5$ years and the threshold of detection as ${\rm SNR}=10$~\cite{Caprini:2015zlo}.

\section{The Shape of the Higgs Potential}
\label{sec:Vshape}

\begin{figure}[!tbp]
    \begin{center}
        \includegraphics[width=0.5\textwidth]{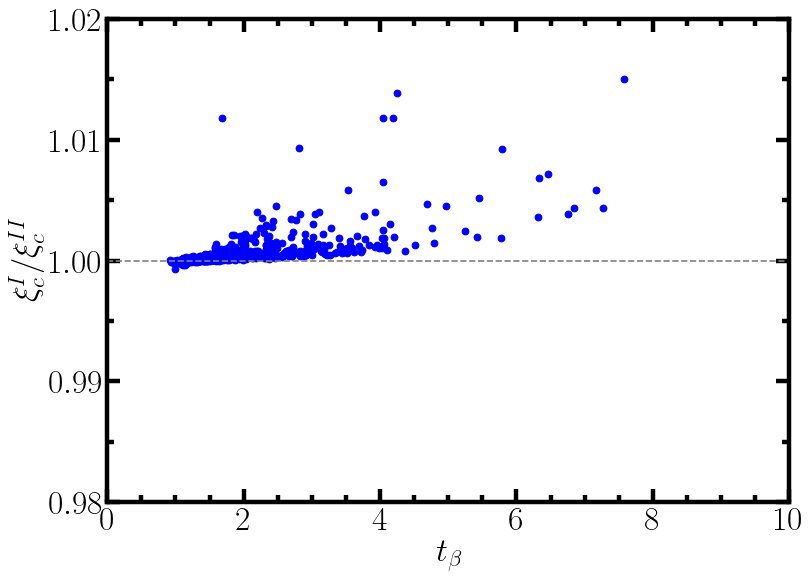}
    \end{center}
    \caption{The ratio of order parameter $\xi_c$ between the type-I and II 2HDM as a function of $t_\beta$. The considered points display strong first-order phase transition $\xi_c>1$. We focus on the points that satisfy all the current constraints for both scenarios. See Sec.~\ref{sec:2HDM} for more details on the parameter space scan. }
    \label{fig:xi_tb_TypeI_TypeII}
\end{figure}

The extension of the SM Higgs sector with another Higgs doublet in the 2HDM can promote the phase transition pattern from a smooth crossover to a strong first-order phase transition.\footnote{In our study, we focus on one-step electroweak symmetry breaking, refraining from addressing multistep phase transitions, which turns out to be far more rare in the 2HDM.}  To study the key ingredients triggering this transmutation in the EWPT, we need to probe the shape and thermal evolution of the effective potential. In this section, we focus on the barrier formation and the upliftment of the true vacuum~\cite{EWPT-NMSSM, EWPT-Nature,Dorsch:2017nza}. The parameter space analysis is organized based on the most relevant components of the effective potential generating these changes in the profile of the Higgs potential~\cite{LianTao:2013cases}.
This characterization is used as a key ingredient to pin down the leading phenomenological parameters for strong first-order phase transition.

In our studies, we focus on the type-I and -II scenarios for the Yukawa couplings. The main difference between these two cases in the effective potential comes from the bottom Yukawa coupling in the Daisy terms.
In terms of phase transition, these two scenarios result in negligible differences.
In~\autoref{fig:xi_tb_TypeI_TypeII}, we show the ratio of $\xi_c$ between type I and II as a function of $t_\beta$ using the same input parameters. We focus on the points that satisfy all the current constraints for both scenarios. We observe depleted differences between $\xi_c^I$ and $\xi_c^{II}$ with most points differing only in the subpercent level. While there is a small enhancement for the ratio $\xi_c^I/\xi_c^{II}$ toward larger $t_\beta$ (as in type II, the Daisy contributions result in a slightly deeper potential at the true vacuum), the difference is phenomenologically insignificant. Since the considered scenarios display a similar phase transition profile, with only subleading differences, we will mostly focus on the type-I case in the present section. Nevertheless, when we discuss the experimental sensitivities, we will show the results for both scenarios as they can present distinct collider phenomenology due to their different fermionic couplings.

\begin{figure}[t]
  \centering
  \includegraphics[width=0.48\textwidth, trim=0 0 100 50]{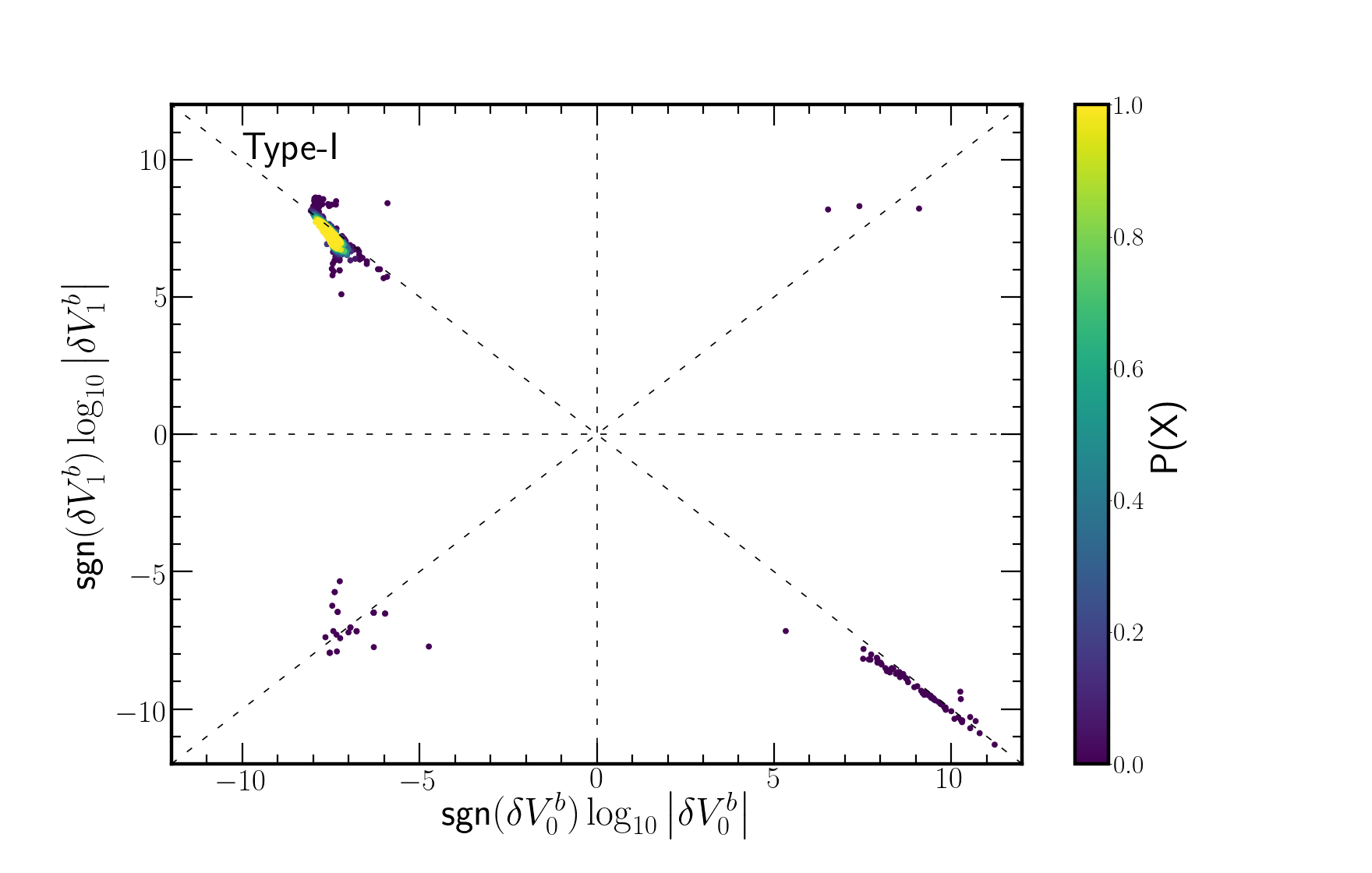}
  \includegraphics[width=0.48\textwidth, trim=0 0 100 50]{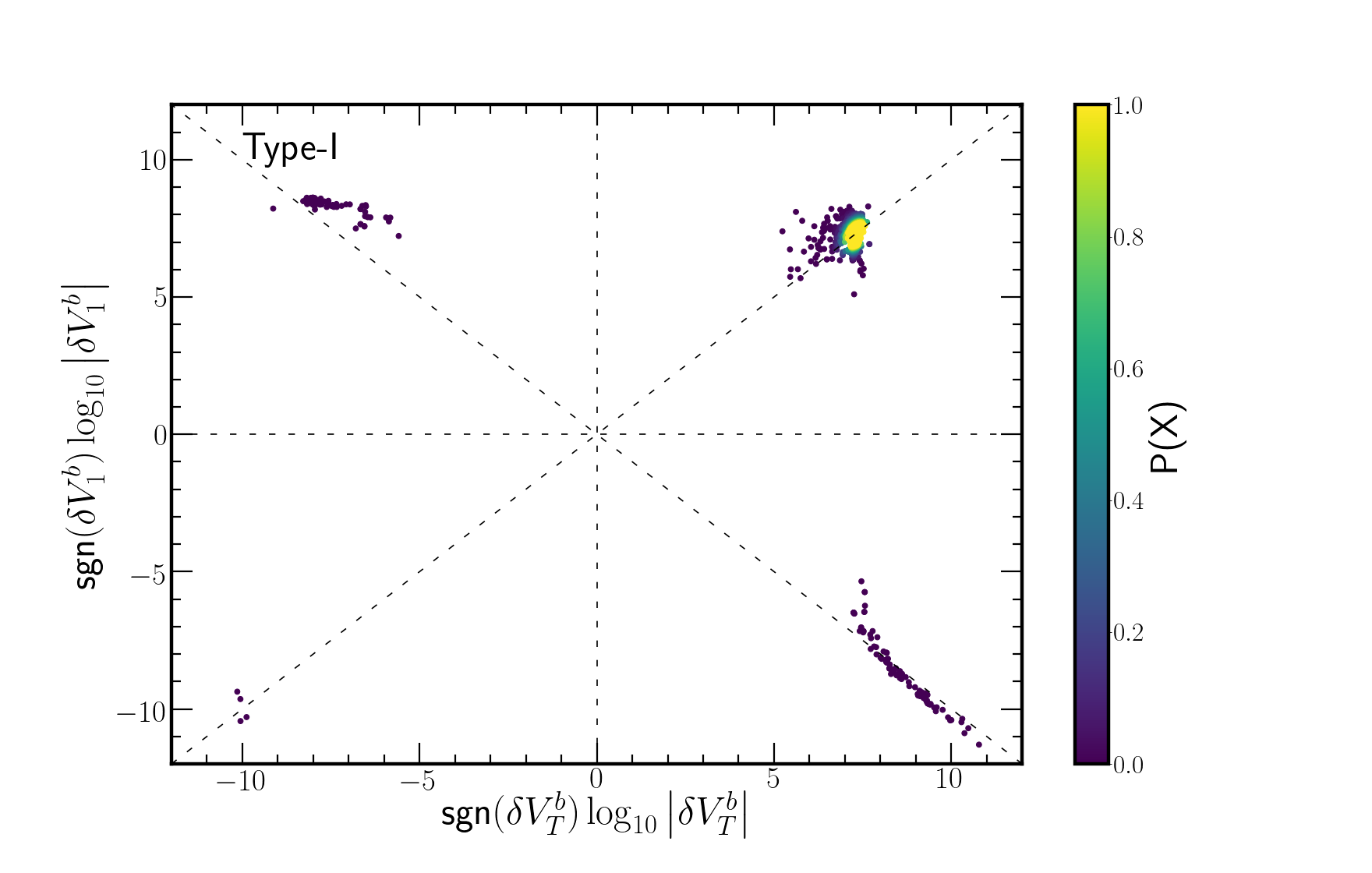}
  \caption{Left panel: contribution of the tree-level potential to the barrier ($V_0^b$) against the one-loop corrections ($V_1^b$).
  Right panel: thermal contributions to the barrier ($V_T^b$) against the one-loop corrections ($V_1^b$).
  The regions are color coded with the probability density of points. We consider the type-I 2HDM requiring $\xi_c>1$.}
  \label{fig:barriercont}
\end{figure}

\subsection{Barrier formation}
\label{sec:barrier}

Moving forward, we scrutinize the 2HDM phase transition pattern, analyzing three classes of contributions to the potential barrier: tree-level ($V_0$), one-loop ($V_1\equiv V_{CW}+V_{CT}$), and thermal effects ($V_T$). Our main target is to identify which of these terms plays a crucial role in introducing the barrier between the broken and unbroken vacua, granting the possibility of strong first-order electroweak phase transition (SFOEWPT) in the 2HDM~\cite{LianTao:2013cases}. The correlations among these contributions to the potential barrier are presented in~\autoref{fig:barriercont}. For illustration, we focus on the type-I 2HDM with $\xi_c>1$. The barrier of the potential is the position where the effective potential obtains the maximum value in the tunneling path obtained by solving \autoref{eq:tunneling}. We defined the height of the barrier as the difference between the effective potential at the barrier and false vacuum. The position of the barrier at $T_c$ is approximated by the point where the potential attains maximum value in the line connecting the true and false vacua. We observe that the potential barrier in the 2HDM is dominantly generated by a coalition between the one-loop and thermal components. These terms display positive contributions to the potential barrier  $\delta V_1^b,\delta V_T^b>0$ for $99\%$ of the parameter space points. In contrast, the tree-level term $V_0$ typically works against the barrier formation.

\begin{figure}[t]
  \centering
    \includegraphics[width=0.49\textwidth,trim=60 50 110 50]{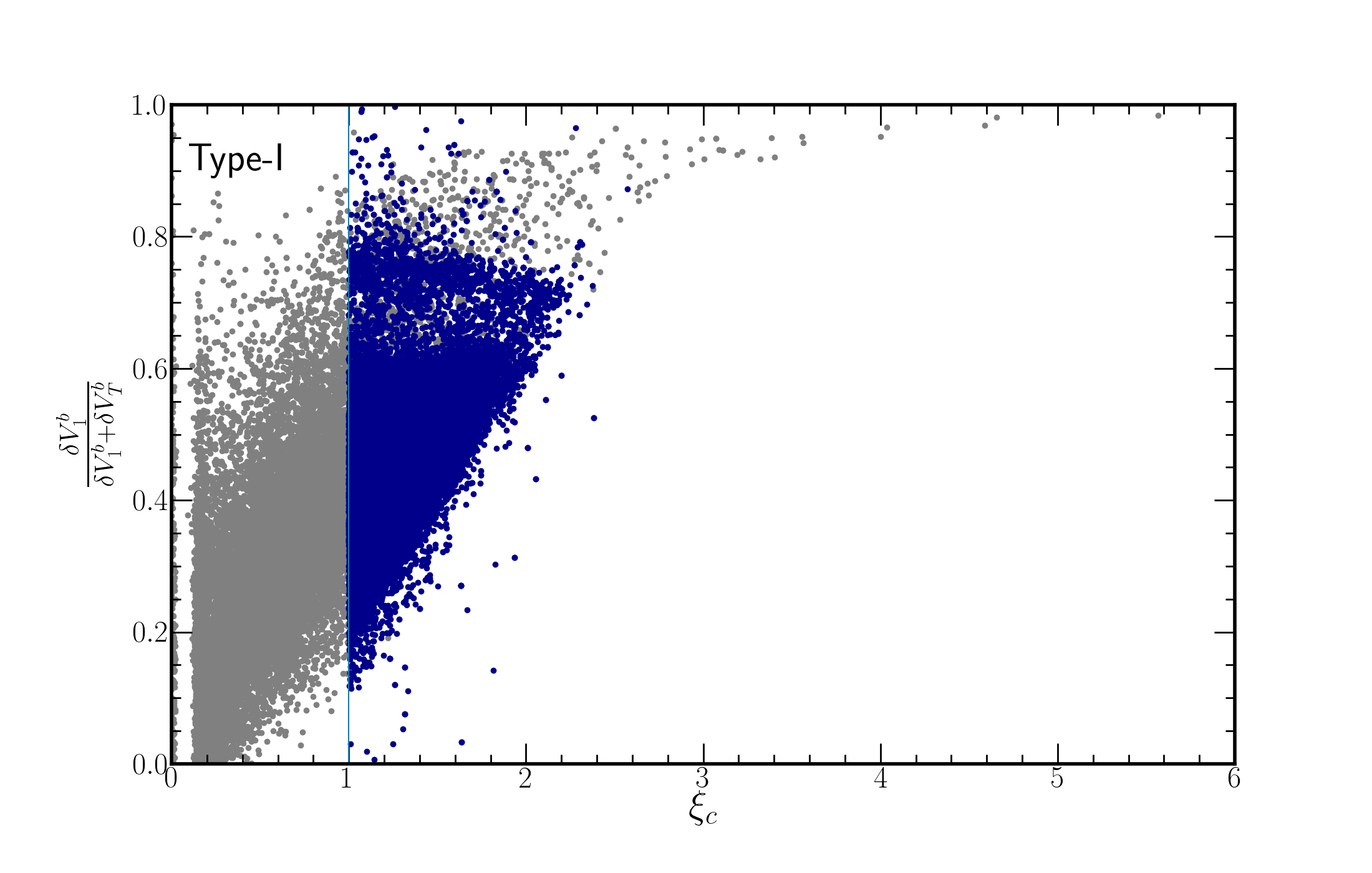}
    \includegraphics[width=0.49\textwidth,trim=60 50 110 50]{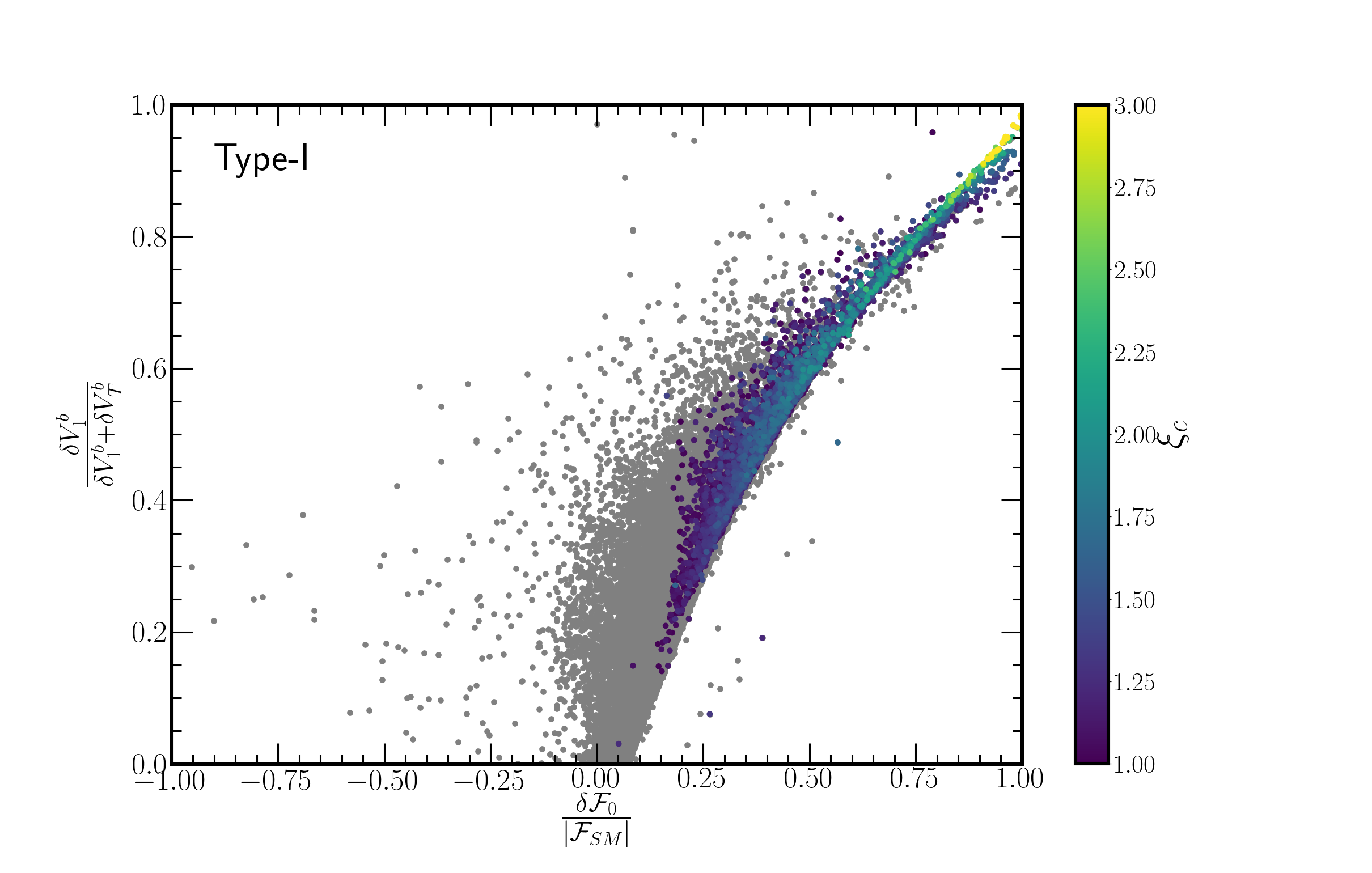}
  \caption{Left panel: the ratio $\frac {\delta V_1^b}{\delta V_1^b+\delta V_T^b}$ for the barrier at $T_c$ versus $\xi_c$. Blue denotes the points with $\xi_c>1$ and have nucleation temperature, while gray represents points with first-order phase transition. Right panel: the ratio $\frac {\delta V_1^b}{\delta V_1^b+\delta V_T^b}$ for the barrier at $T_c$ versus $\delta \mathcal{F}_0/\mathcal{F}_0^{\rm SM}$ color coded with $\xi_c$. Gray denotes all first-order phase transition points. We assume the type-I 2HDM, focusing on the most probable region in~\autoref{fig:barriercont}, where the barrier is generated by the one-loop and thermal corrections $\delta V_1^b,\delta V_T^b>0$.}
  \label{fig:barrierratio}
\end{figure}

 In~\autoref{fig:barrierratio} (left panel), we present the fraction between the two leading terms to the potential barrier as a function of the order parameter $\xi_c$. We focus on the region with positive contributions  $\delta V_1^b,\delta V_T^b>0$ enclosing the bulk of the parameter space points. Two comments are in order. First, in the strong first-order phase transition regime $\xi_c>1$, the phase transition is mostly one-loop driven; i.e., the effective potential barrier is dominantly generated by the one-loop term. In this case, the $\hbar$ loop corrections can generate relevant nonpolynomial field dependencies, such as $h^4\ln h^2$, that contribute to the barrier formation~\cite{LianTao:2013cases,Reichert:2017puo}. Second, if the fraction of the barrier height provided by the one-loop contribution is close to 100\%, the tunneling from the false vacuum (metastable vacuum) to the true vacuum is more challenging. For this reason, the universe with $\xi_c\gtrsim 2.5$ is trapped in the false vacuum
 and electroweak symmetry breaking does not occur. We should notice that this feature is associated with the dominant phase space regime $\delta V_1^b,\delta V_T^b>0$. Conversely, the rarer tree-level or thermally driven setups can still generate stronger phase transitions $\xi_c\gtrsim 2.5$.

\subsection{Vacuum upliftment}
\label{sec:vauum-uplift}

After looking at the general new physics contributions producing the barrier, we now focus on the effects on the potential at the vacua. The strength of the phase transition has been shown to be correlated with the upliftment of the true vacuum compared to the symmetric one at zero temperature~\cite{EWPT-NMSSM, EWPT-Nature,Dorsch:2017nza}. That is, if the Higgs potential is shallow at $T=0$, the required thermal upliftment for SFOEWPT, making the true vacuum degenerate with the false one, is reduced. Following a similar notation to Ref.~\cite{Dorsch:2017nza}, we define a dimensionless parameter to measure the true vacuum upliftment
\begin{equation}
    \frac {\Delta \mathcal{F}_0}{|\mathcal{F}_0^{\rm SM}|} \equiv \frac {\mathcal {F}_0-\mathcal{F}_0^{\rm SM}}{|\mathcal{F}_0^{\rm SM}|},
\end{equation}
where $\mathcal{F}_0$ is the  vacuum energy density of the 2HDM at $T=0$  defined as
\begin{equation}
    \mathcal{F}_0\equiv V_{\rm eff}(v_1,v_2,T=0)-V_{\rm eff}(0,0,T=0),
\end{equation}
and $\mathcal{F}_0^{\rm SM}=-1.25 \times 10^8 ~\text{GeV}^4$.  In~\autoref{fig:barrierratio} (right panel), we note that the barrier height provided by the one-loop contribution is correlated with $\Delta \mathcal{F}_0/ | \mathcal{F}_0^{\rm SM}|$, which measures the vacuum upliftment at zero temperature. The larger the one-loop contribution, the higher the vacuum upliftment. This correlation is especially prominent for $\xi_c>1$.
Since the one-loop effects are dominant with respect to thermal corrections for $\xi_c>1$,  $\Delta \mathcal{F}_0/ | \mathcal{F}_0^{\rm SM} |$ works as a good first approximation to study some general properties of the EWPT, even though it is a zero-temperature quantity. In particular, it is possible to define a typically necessary condition for a first-order phase transition with the minimum threshold $\Delta \mathcal{F}_0/ | \mathcal{F}_0^{\rm SM} |>0.34$~\cite{Dorsch:2017nza}. Although this condition encapsulates most of the $\xi_c>1$ points, we should note from~\autoref{fig:barrierratio} that it does not work as a sufficient requirement for SFOEWPT.

\begin{figure}
	\centering
	\includegraphics[width=0.5\textwidth]{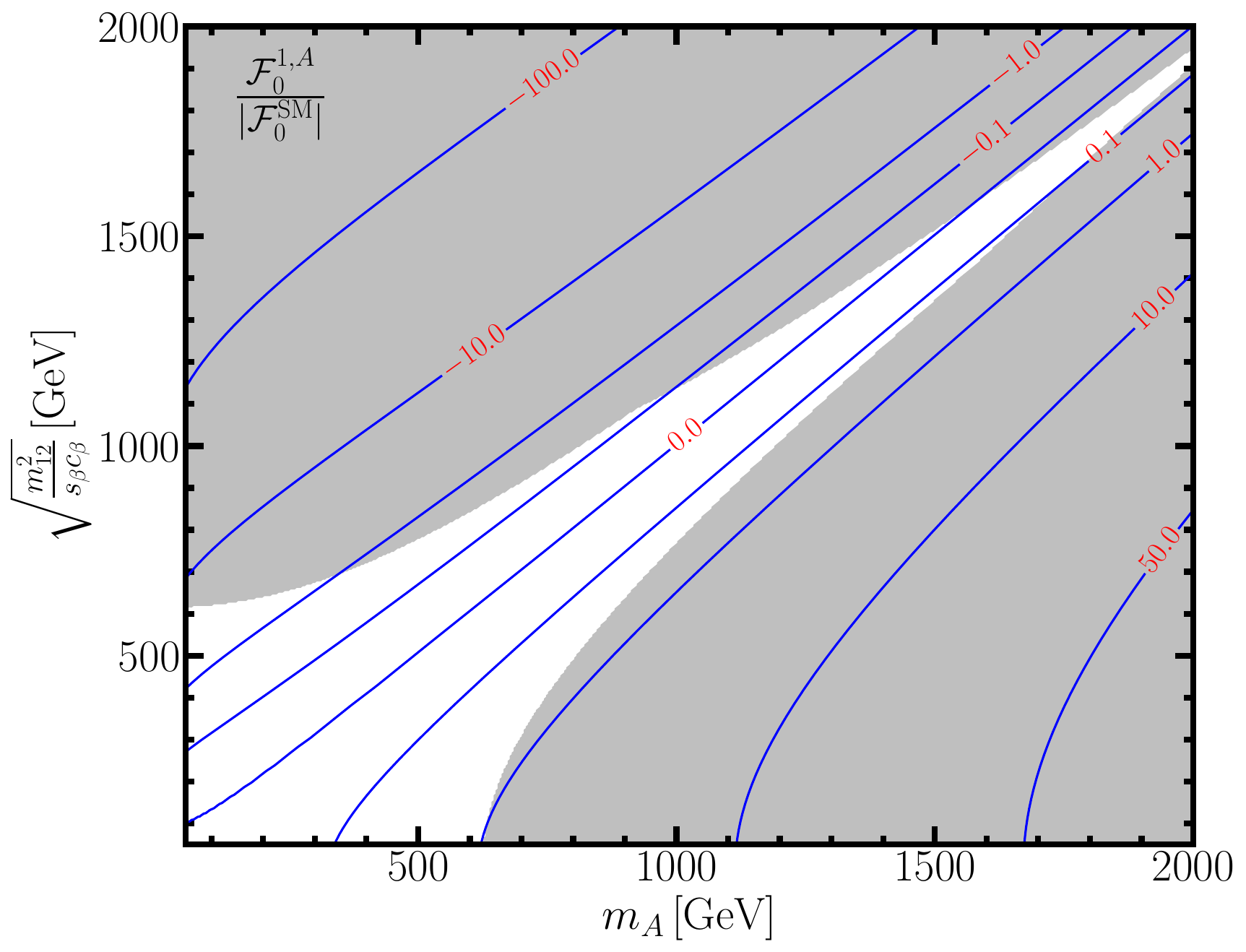}
	\caption{The individual contribution  $\mathcal{F}_0^{1,A}/|\mathcal{F}_0^{\rm SM}|$ from $A$ (blue solid line) in the $m_A$-$\sqrt{\frac{m_{12}^2}{s_\beta c_\beta}}$ plane. The contributions from $H$ and $H^\pm$ have the same form. The number in red inside each line indicates the value for $\mathcal{F}_0^{1,A}/|\mathcal{F}_0^{\rm SM}|$. We assume the alignment limit. The gray shaded region is excluded by unitarity and perturbativity constraints assuming $t_\beta = 1$, $c_{\beta-\alpha}=0$, and $m_A=m_{H^\pm}=m_H+100$ GeV.}
	\label{fig:dF0_individual}
\end{figure}

While $\Delta \mathcal{F}_0/| \mathcal{F}_0^{\rm SM} |$ does not provide a one-to-one correlation with the strength of EWPT, it is helpful to explain critical features of the 2HDM parameter space. In the alignment limit, the important contributions to $\mathcal{F}_0$ come from $H,~A,$ and $H^\pm$ which can be written as
\begin{align}
    \mathcal{F}_0^{1,H}(c_{\beta-\alpha}=0) &= \frac{1}{512\pi^2}\left[\left(3m_h^2+2m_H^2-6\frac{m_{12}^2}{s_\beta c_\beta}\right)\left(m_h^2+2m_H^2-2\frac{m_{12}^2}{s_\beta c_\beta}\right)\right. \nonumber\\
    &\qquad\qquad\qquad \left.+ \left(m_h^2-2 \frac{m_{12}^2}{s_\beta c_\beta}\right)^2\log \left(\frac{4m_H^4}{(m_h^2-2 m_{12}^2/(s_\beta c_\beta))^2}\right)\right]\,, \nonumber\\
    \mathcal{F}_0^{1,A}(c_{\beta-\alpha}=0) &= \frac{1}{512\pi^2}\left[\left(3m_h^2+2m_A^2-6\frac{m_{12}^2}{s_\beta c_\beta}\right)\left(m_h^2+2m_A^2-2\frac{m_{12}^2}{s_\beta c_\beta}\right)\right. \nonumber\\
    &\qquad\qquad\qquad \left.+ \left(m_h^2-2 \frac{m_{12}^2}{s_\beta c_\beta}\right)^2\log \left(\frac{4m_A^4}{(m_h^2-2 m_{12}^2/(s_\beta c_\beta))^2}\right)\right]\,,  \nonumber\\
    \mathcal{F}_0^{1,H^\pm}(c_{\beta-\alpha}=0) &= \frac{1}{256\pi^2}\left[\left(3m_h^2+2m_{H^\pm}^2-6\frac{m_{12}^2}{s_\beta c_\beta}\right)\left(m_h^2+2m_{H^\pm}^2-2\frac{m_{12}^2}{s_\beta c_\beta}\right)\right. \nonumber\\
    &\qquad\qquad\qquad \left.+ \left(m_h^2-2 \frac{m_{12}^2}{s_\beta c_\beta}\right)^2\log \left(\frac{4m_{H^\pm}^4}{(m_h^2-2 m_{12}^2/(s_\beta c_\beta))^2}\right)\right]\,.
    \label{eq:dF0exp}
\end{align}
Using these expressions, we can write the shift in vacuum energy density with respect to the SM value with
\begin{align}
\Delta \mathcal{F}_0=\mathcal{F}_0-\mathcal{F}_0^{\rm SM}=\mathcal{F}_0^{1,H}+\mathcal{F}_0^{1,A}+\mathcal{F}_0^{1,H^\pm}\,.
\end{align}
Remarkably, the individual contributions from $H$, $A$, and $H^\pm$ to $\mathcal{F}_0$ are of the same form. Thus, in this sense, there should be no difference in the preferred region in terms of $m_H$, $m_A$, and $m_{H^\pm}$. On the other hand, $m_{12}^2/(s_\beta c_\beta)$ also plays an important role. In~\autoref{fig:dF0_individual}, the individual contributions from $A$ to $\mathcal{F}_0$ are shown ($H$ and $H^\pm$ have the same form), from which it is easy to find that $\mathcal{F}_0^{1,\Phi}$ will be negative when ${m_{12}/\sqrt{s_\beta c_\beta}}$ is larger than the scalar mass $m_\Phi$, where $\Phi=H,~A$, or $H^\pm$. The larger the difference, the more negative it will be. Contrarily, when the scalar mass is larger than ${m_{12}/\sqrt{s_\beta c_\beta}}$, $\mathcal{F}_0$ tends to be positive, uplifting the true vacuum, and favoring the strong first-order phase transition.
However, the vacuum upliftment $\Delta\mathcal{F}_0$ is limited from above and below by perturbative unitarity constraints~\cite{Ginzburg:2005dt}.
For illustration, as a benchmark, we denote the {perturbative unitarity} constraints as the gray shaded region in~\autoref{fig:dF0_individual} for $t_\beta = 1$, $c_{\beta-\alpha}=0$, and $m_A=m_{H^\pm}=m_H+100$ GeV. As the allowed region (unshaded area) narrows toward larger scalar masses and $m_{12}/\sqrt{s_\beta c_\beta}$, sizable scalar masses admit only a small vacuum upliftment $\Delta\mathcal{F}_0$. Therefore, SFOEWPT generally favors low scalar masses, granting larger (and positive) $\Delta\mathcal{F}_0$. These analytical results are in accordance with the rather general arguments from Ref.~\cite{Ramsey-Musolf:2019lsf}, in which the author also argued that light scalars are favored.

\begin{figure}[!tbp]
    \centering
     \includegraphics[width=\textwidth]{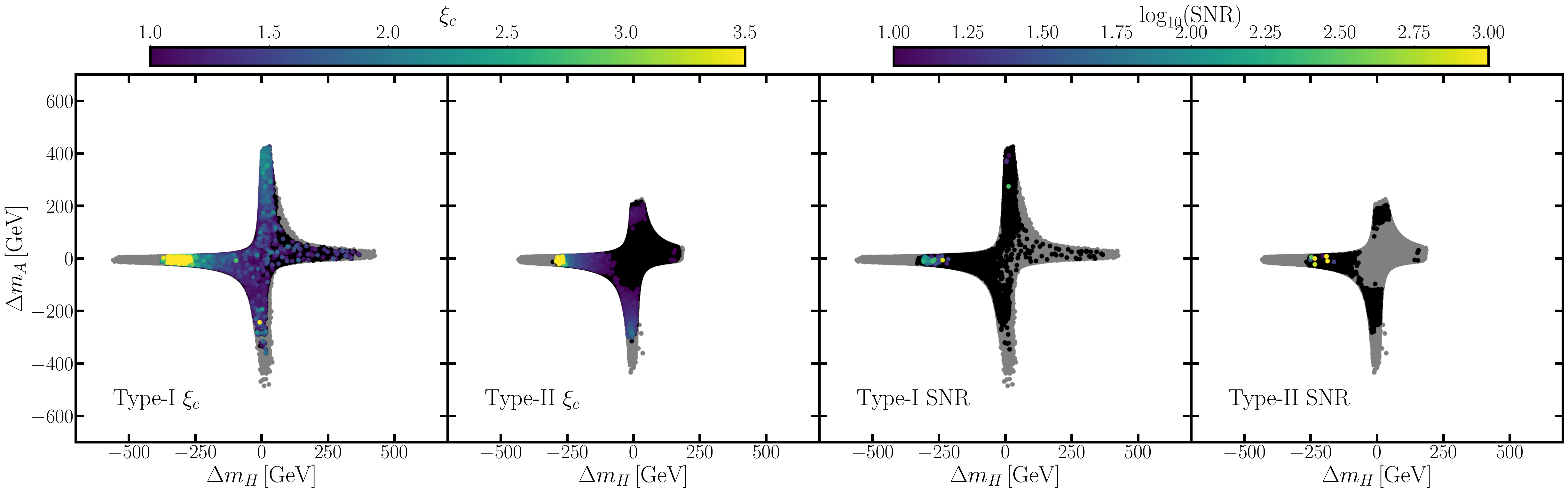}
      \caption{Parameter space scan in terms of $(\Delta m_H,\Delta m_A)$. The heat map tracks $\xi_c$ (left two panels) and the SNR (right two panels). For the study in terms of the order parameter $\xi_c$ (left two panels), the black points represent the parameter space regime with first-order phase transition with $\xi_c>0$. In contrast, for the SNR analysis (right two panels), the black points characterize  the regime with $\xi_c>1$. The parameter space scan is performed with {\tt ScannerS}~\cite{Muhlleitner:2020wwk}, where we impose the  constraints from perturbative unitarity, boundedness from below, vacuum stability, electroweak precision, and flavor constraints. {\tt HiggsBounds} and {\tt HiggsSignals} are used to incorporate the searches for additional scalars as well as the 125~GeV Higgs boson measurements~\cite{Bechtle:2020pkv,Bechtle:2020uwn}. For more details on the parameter space scan, see~\autoref{sec:2HDM}.}
     \label{fig:dma_dmh_xi_SNR}
\end{figure}

It is possible to shed more light on the general profile of EWPT in the 2HDM, combining the $\mathcal{F}_0$ dependence on the new scalar masses with the constraints from electroweak precision measurements, which require $m_H$ or $m_A$ to be close to $m_{H^\pm}$. In~\autoref{fig:dma_dmh_xi_SNR}, we show the scanned points in the $(\Delta m_H,\Delta m_A)$ plane, where $\Delta m_H \equiv m_H - m_{H^\pm}$ and $\Delta m_A \equiv m_A - m_{H^\pm}$. The heat map tracks $\xi_c$ (left two panels) and the SNR (right two panels).\footnote{We thank the authors of Ref.~\cite{Moretti:2026qsw} for pointing out an issue with the SNR values reported in the previous version of the manuscript, which has been corrected in the present version. The correction modifies the quantitative LISA sensitivity, while leaving unchanged the qualitative hierarchy found in the original analysis: the HL-LHC probes a substantially larger fraction of the SFOEWPT parameter space than LISA.} For the study in terms of the order parameter $\xi_c$ (left panels), the black points represent the parameter space regime with first-order phase transition with $\xi_c>0$. In contrast, for the SNR analysis (right panels), the black points characterize the regime with $\xi_c>1$. The gray points in the background pass all the theoretical and current experimental constraints. The results display a general cross pattern in the $(\Delta m_H,\Delta m_A)$ plane dominantly induced by the electroweak precision measurements~\cite{Grimus:2007if,Gerard:2007kn}. There are significant differences between the type-I and -II scenarios, prompted by the flavor physics constraints. More concretely, bounds from B-meson decays require $m_{H^\pm}\gtrsim580$ GeV in type-II 2HDM. In particular, this lower bound on the charged scalar mass cuts off part of the top and right $(\Delta m_H,\Delta m_A)$ branches compared with the type-I scenario. The type-I and -II 2HDMs present a similar phase transition profile for the remaining parameter space points, as shown in \autoref{fig:xi_tb_TypeI_TypeII}.

An immediate feature of the right panels of~\autoref{fig:dma_dmh_xi_SNR} is that LISA is sensitive to only a limited subset of the $\xi_c>1$ parameter space. Indeed, $\xi_c>1$ alone does not guarantee an observable gravitational wave signal, whose strength is instead governed by the phase transition dynamics. In particular, sizable signals require sufficiently large latent heat $\alpha$ and a sufficiently slow transition, corresponding to small $\beta/H_n$. These conditions are satisfied only for a restricted fraction of the SFOEWPT points.

Before discussing the phase transition pattern presented in~\autoref{fig:dma_dmh_xi_SNR}, we would like to point out several theoretical constraints that will be important for our analysis. Especially, we want to highlight that $m_{12}/\sqrt{s_\beta c_\beta}$ cannot be too different from the scalar masses. First, we consider the perturbative constraints. We start by writing the $\lambda_1$ and  $\lambda_2$ couplings in the alignment limit
\begin{align}
    \lambda_1 v^2 &\approx m_h^2 + t_\beta^2 \left(m_H^2 - \frac{m_{12}^2}{s_\beta c_\beta}\right)\,,\nonumber \\
    \lambda_2 v^2 &\approx m_h^2 + \frac{1}{t_\beta^2} \left(m_H^2 - \frac{m_{12}^2}{s_\beta c_\beta}\right)\,.
    \label{eq:lamb1}
\end{align}
Because of their strong $t_\beta$ dependence, perturbativity limits for $\lambda_1$ and  $\lambda_2$ demand $m_H^2 \approx  m^2_{12}/(s_{\beta}c_{\beta})$ for $t_\beta$ significantly different from 1. Second, the boundedness from below limits~\cite{Ivanov:2018jmz} ${\lambda_{1,2}>0}$ requires that $m_{12}^2/(s_\beta c_\beta)$ cannot be much larger than $m_H^2$.

Exploring the aforementioned theoretical constraints and the scalar contributions to $\mathcal{F}_0$, we scrutinize the general phase transition profile shown in~\autoref{fig:dma_dmh_xi_SNR} by looking at each of the  $(\Delta m_H,\Delta m_A)$ branches.
\begin{itemize}
    \item $m_H<m_{H^\pm}\approx m_A$ (left branch): This mass configuration displays numerous first-order phase transition points.
   Since $m_{12}^2/(s_\beta c_\beta)$ cannot be significantly larger than $m_H^2$, this leads to large and positive
    contributions from $A$ and $H^\pm$ to $\mathcal{F}_0$  associated with typically small effects from $H$. These properties promote the $m_H<m_{H^\pm}\approx m_A$ regime as one of the most likely configurations to achieve SFOEWPT in the 2HDM. Further, when the mass difference $|\Delta m_H|$ is large, the contributions to $\mathcal{F}_0$ from $H^\pm$ and $A$ are also sizable. Hence, large $|\Delta m_H|$ is more likely to grant SFOEWPT.\footnote{However, there is  an upper bound on $|\Delta m_H|$. The electroweak symmetry breaking vacuum becomes metastable when the contribution to $\mathcal{F}_0$ is too sizable.} Remarkably, this results in important phenomenological consequences for LHC searches. In particular, it favors new physics searches via the $A\to ZH$ channel that will be discussed in Sec.~\ref{sec:AZH}.
    \item $m_H\approx m_{H^\pm}<m_A$ (top branch): This regime also presents a sizable number of $\xi_c>1$ points. As $m_{12}^2/(s_\beta c_\beta)$ cannot be much larger than $m_H^2$, the leading positive contribution to $\mathcal{F}_0$ arises from the pseudoscalar $A$. In addition, when $m_{12}^2/(s_\beta c_\beta)$ is smaller than $m_{H(H^\pm)}^2$, $H$ $(H^\pm)$ can also contribute positively, while at a subleading level when compared to $A$. Hence, the order parameter $\xi_c$ tends to be slightly suppressed, in comparison to the left branch. {Finally, due to similar arguments as for the left branch, sizable $|\Delta m_A|$ is more likely to yield SFOEWPT.}
    \item $m_H\approx m_{H^\pm} > m_A$ (bottom branch): This region can provide SFOEWPT, as long as $m_{12}^2/(s_\beta c_\beta)$ is much lower than $m_H^2\approx m_{H^\pm}^2$. In this regime, the constraints from $\lambda_{1,2}$ imply $t_\beta\to 1$ to generate a large order-parameter, $\xi_c>1$.
    \item $m_H>m_{H^\pm}\approx m_A$ (right branch): This mass configuration renders a suppressed number of first-order phase transition points. Similar to the bottom branch, $m_{12}^2/s_\beta c_\beta$ has to be much lower than $m_H^2$ to achieve the $\xi_c>1$ regime.
    However, in this parameter space, only $H$ can contribute significantly to $\mathcal{F}_0$, which leads to a lower chance to achieve SFOEWPT.
    \item $m_H\approx m_{H^\pm} \approx m_A$ (central region): This region has a depleted number of SFOEWPT points. All the considered masses are close to each other, as well as $m_{12}/\sqrt{s_\beta c_\beta}$. Thus, all their contributions to $\mathcal{F}_0$ will be suppressed. Notably, due to the charged Higgs mass constraint $m_{H^\pm}>580$~GeV, the type-II 2HDM displays further suppression on the number of $\xi_c>1$ points for this region in comparison to the type I.
\end{itemize}

Combining all these arguments, we find that (i) $m_H<m_{H^\pm}\approx m_A$ provides the most likely regime to accommodate first-order EWPT, (ii) $m_H\approx m_{H^\pm}<m_A$, followed by $m_H\approx m_{H^\pm}>m_A$, also have a high probability of producing $\xi_c>1$, and (iii) it is more likely to have SFOEWPT for larger mass differences $\Delta m_{H,A}$. Thus, the degenerate mass spectrum $m_H\approx m_{H^\pm} \approx m_A$ depletes the SFOEWPT points. However, when the mass difference is exceedingly large, as in the left and bottom branches, it renders the EW vacuum unstable~\cite{Ivanov:2015nea}.

\section{Collider and Gravitational Wave Signals}
\label{sec:Collider_GW}
\begin{figure}[!tbp]
    \centering
     \includegraphics[width=\textwidth]{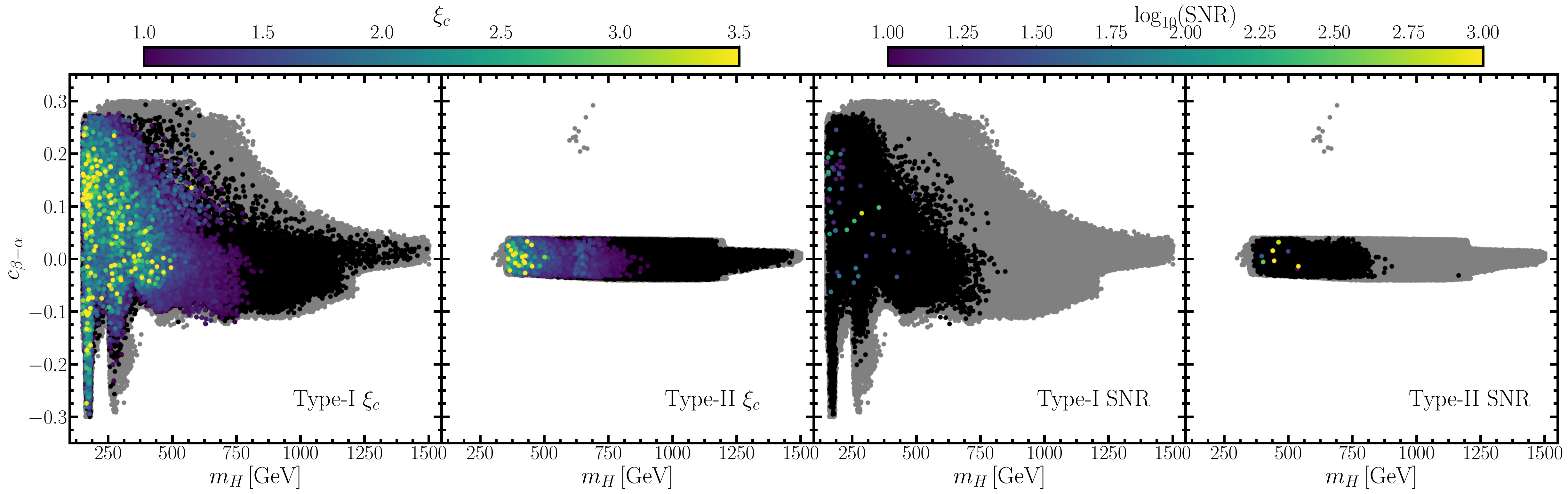}
     \caption{Scan points in the $m_H$-$c_{\beta-\alpha}$ plane. The color code is the same as~\autoref{fig:dma_dmh_xi_SNR}.
	 }
     \label{fig:cba_MHH_xi_SNR}
 \end{figure}

In this section, we focus on the complementarities between collider and gravitational wave experiments to probe the phase transition pattern in the early Universe. In~\autoref{fig:cba_MHH_xi_SNR}, we show the scanned points in  the $(m_H,c_{\beta-\alpha})$ plane. The current experimental constraints restrict the 2HDM toward the alignment limit, $c_{\beta-\alpha}\to 0$.\footnote{While most experimental limits are generally symmetric with respect to $c_{\beta-\alpha}=0$, the searches that involve, in particular, the $Hhh$ interaction do not satisfy this form~\cite{ATLAS:2019qdc}. This asymmetry is more prominent for the type I in the $m_H\gtrsim 2 m_{h}$ regime, where the resonant double Higgs production $H\to hh$ is kinematically allowed.} Although it is possible to have arbitrarily large new scalar masses in the 2HDM, SFOEWPT and GW observability generally limit these new scalar modes to below the TeV scale~\cite{Ramsey-Musolf:2019lsf}. In~\autoref{fig:cba_MHH_xi_SNR} (left panel), the $\xi_c>1$ condition results in a typical upper limit on the heavy scalar mass of ${m_H\lesssim 750}$~GeV. The lighter the resonance, the higher the order parameter. As shown in~\autoref{sec:Vshape}, this can be explained by an analysis of the theoretically allowed range for the new physics contributions to $\Delta\mathcal{F}_0$. While sizable scalar masses  grant only a small vacuum upliftment $\Delta\mathcal{F}_0$, modest scalar masses can display large (and positive) $\Delta \mathcal{F}_0$. This renders a strong extra motivation for scalar searches, in the 2HDM, at the high-luminosity LHC. Hence, in this section, along with the theoretical and current experimental limits, we will discuss the HL-LHC projections of relevant 2HDM searches and contrast them with the sensitivity to SFOEWPT and GW observation. In the following, the relevant cross sections are obtained from {\tt ScannerS}~\cite{Muhlleitner:2020wwk}. It includes a tabulated parameterization of the next-to-next-to-leading-order (NNLO) QCD gluon fusion and $bb$-associated Higgs production obtained from SusHi~\cite{Harlander:2012pb,Harlander:2016hcx}. It also encompasses the  next-to-leading-order (NLO) QCD top quark and charged Higgs boson associated production parametrized within HiggsBounds~\cite{Bechtle:2020pkv,Berger:2003sm,Dittmaier:2009np,Flechl:2014wfa,Degrande:2015vpa,LHCHiggsCrossSectionWorkingGroup:2016ypw,Degrande:2016hyf}.

\subsection{Resonant and nonResonant di-Higgs searches}
\label{sec:hh}

\begin{figure}[!tbp]
    \begin{center}
        \includegraphics[width=\textwidth]{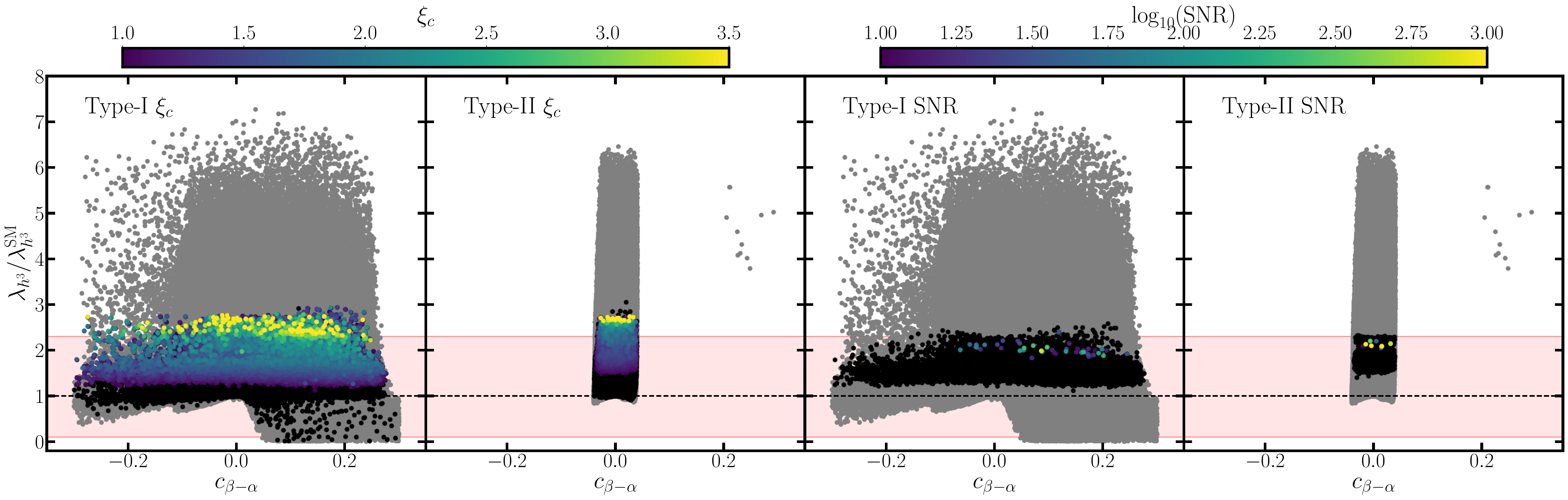}
    \end{center}
    \caption{The triple Higgs coupling normalized to the SM value as a function of $c_{\beta-\alpha}$. The color code is the same as~\autoref{fig:dma_dmh_xi_SNR}. The HL-LHC projected 95\% C.L. sensitivity for nonresonant di-Higgs production is also shown $0.1<\lambda_{h^3}/\lambda_{h^3}^{\rm SM}<2.3$ (light red)~\cite{Cepeda:2019klc}.
    }
    \label{fig:CHHH_cba_xi_SNR}
\end{figure}

While in the alignment limit, the tree-level Higgs self-coupling matches the SM value  $\lambda_{h^3}=\lambda_{h^3}^{\rm SM}$, the one-loop corrections in the 2HDM can significantly disrupt this equality. In~\autoref{fig:CHHH_cba_xi_SNR}, we show the new physics effects on the triple Higgs coupling $\lambda_{h^3}/\lambda_{h^3}^{\rm SM}$ at one loop. We observe that the higher-order effects can produce extremely high deviations from the SM, as large as $\lambda_{h^3}/\lambda_{h^3}^{\rm SM}\approx 7$, even in the alignment limit and in view of the theoretical and experimental constraints. Remarkably, the sizable radiative corrections do not translate in the breakdown of validity for the perturbation theory. Instead, they are a result of new one-loop contributions coming from other types of couplings, such as $\lambda_{hHH}$ and $\lambda_{hhHH}$. The corrections are naturally expected to stabilize beyond one loop, where these new types of effects are already accounted for~\cite{Kanemura:2002vm}.

The new physics contributions to triple Higgs couplings are occasionally expressed in the effective field theory (EFT) framework, integrating out the heavy modes.
Although this approach presents a systematic pathway to include new physics effects in terms of an expansion associated with the new physics scale, it does not generally warrant an appropriate description for the underlying beyond-Standard-Model (BSM) physics, as it  strongly depends on the decoupling of the heavy states~\cite{Appelquist:1974tg}.  In view of the preference for relatively light scalar modes ${m_H\lesssim 750}$~GeV, producing sizable order parameter $\xi_c>1$, the EFT does not provide a robust framework to study the preferred SFOEWPT parameter space regime for the 2HDM at the LHC~\cite{Postma:2020toi}.
Hence, as our limits strongly depend on the LHC results, we refrain from using the EFT approach in this study.

The Higgs pair production $pp\to hh$ provides a direct probe for the Higgs self-coupling at colliders~\cite{EBOLI1987269,Plehn:1996wb,Dolan:2012rv}. However, the limited production rate, large destructive interference between the triangle and box diagrams, and sizable backgrounds make this analysis extremely challenging at the LHC. The ATLAS and CMS high-luminosity projections constrain the triple Higgs coupling at 95\% C.L. to~\cite{Cepeda:2019klc}
\begin{align}
 0.1<\lambda_{h^3}/\lambda_{h^3}^{\rm SM}<2.3   \,.
\end{align}
This limited precision prompts the Higgs self-coupling as a key benchmark for future colliders. In particular, the rapid increase of the gluon luminosity at higher energies translates in a  sizable $pp\to hh$ cross section at the $100$~TeV Future Circular Collider. In such a setup, the large number of signal events would transform the di-Higgs production into a precision measurement, allowing for the full kinematic exploration, that is central for a better resolution of $\lambda_{h^3}$~\cite{Goncalves:2018qas}. The improvement to the Higgs self-coupling measurement at higher energy colliders would allow for a more meaningful global fit analysis that is sensitive to a more complete set of new physics modifications to the Higgs potential~\cite{Biekotter:2018jzu}. Despite the limited experimental constraints, we observe in~\autoref{fig:CHHH_cba_xi_SNR} that the HL-LHC will be sensitive to a large range of the 2HDM parameter space. In particular, it will be able to probe a substantial fraction of points with a large order-parameter that generally correlates with sizable triple Higgs coupling.

While the HL-LHC will be mostly sensitive to SFOEWPT with large order-parameter $\xi_c\gtrsim 2.5$, LISA will be sensitive to GW signals in the complementary regime $\xi_c\lesssim 2.5$, as shown by the color points of~\autoref{fig:CHHH_cba_xi_SNR} (right two panels). As explained in~\autoref{sec:Vshape}, extremely large $\xi_c$ leads to the configuration where the system is trapped in the false vacuum precluding successful nucleation.
This is  reflected in~\autoref{fig:CHHH_cba_xi_SNR}, where the points with large $\lambda_{h^3}$ (and thus large $\xi_c$) in the left two panels do not appear in the right two panels.

\begin{figure}[!tb]
    \begin{center}
        \includegraphics[width=\textwidth]{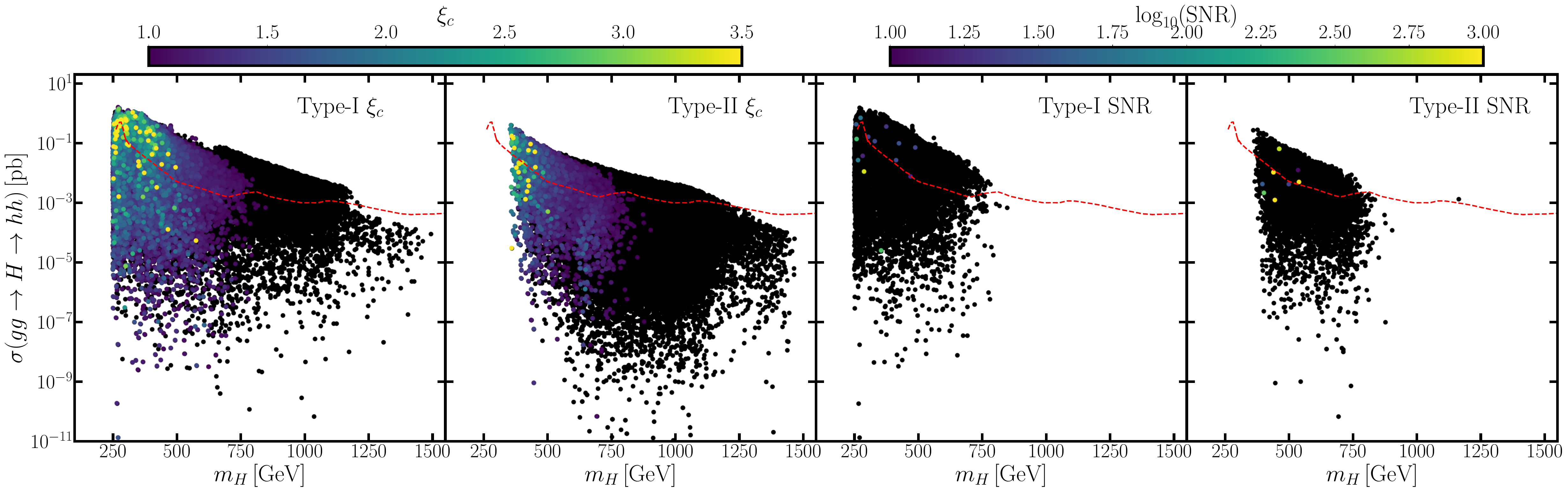}
    \end{center}
    \caption{The cross section $\sigma(gg\to H)\times{\mathcal{BR}}(H\to hh)$ vs. $m_H$. The red dashed line indicates the projected limits from ATLAS with $3\,{\rm ab}^{-1}$ by scaling the current limits from Ref.~\cite{Aaboud:2018knk}. The color code is the same as~\autoref{fig:dma_dmh_xi_SNR}. }
    \label{fig:CS_MHH_xi_SNR}
\end{figure}

Resonant di-Higgs searches provide another prominent probe for the phase transition pattern in the early Universe.
As discussed in the last section,
SFOEWPT is usually associated with light extra scalars $m_H\lesssim 750$~GeV resulting in a favored energy range for $pp\to H\to hh$ production at the LHC.
In~\autoref{fig:CS_MHH_xi_SNR}, in addition to the current theoretical and experimental limits, we present the projected HL-LHC sensitivity at 95\% C.L. to the resonant di-Higgs cross section (red dashed line). The results are obtained scaling the current sensitivity presented by ATLAS in Ref.~\cite{Aaboud:2018knk}, according to the luminosity, to the high-luminosity LHC with $\mathcal{L}=3~\text{ab}^{-1}$. This experimental study focuses on the leading $4b$ final state channel. While this analysis explores the gluon fusion production mode, we note that the resonant production through weak boson fusion can provide additional relevant extra sensitivity~\cite{Barman:2020ulr}. We observe in~\autoref{fig:CS_MHH_xi_SNR} that the projected resonant Higgs pair production measurement will be able to cover a significant part of the parameter space with large $\xi_c$. In contrast, LISA will be sensitive to only a small portion of the parameter space probed by the HL-LHC, as well as to a small complementary subset of points for which the production cross section $pp\to H\to hh$ is suppressed.

\subsection{$A\to ZH$ and $H\to ZA$ searches}
\label{sec:AZH}

\begin{figure}[!tbp]
    \centering
    \includegraphics[width=\textwidth,trim=0 75 0 0,clip]{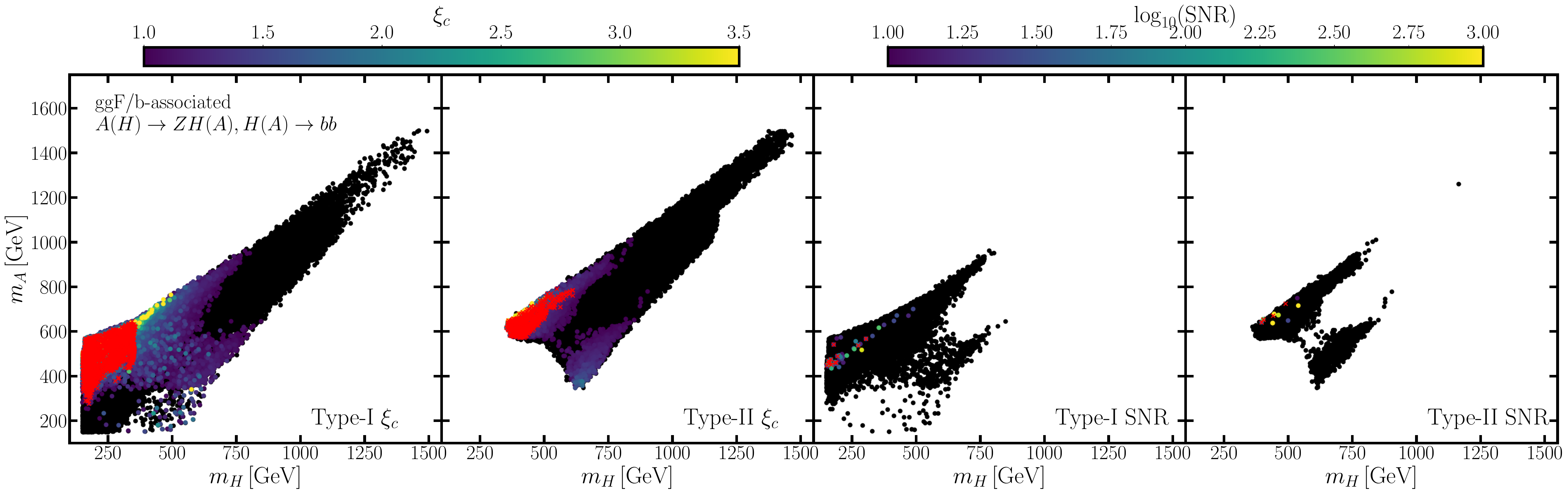}
    \includegraphics[width=\textwidth,trim=0 0 0 110,clip]{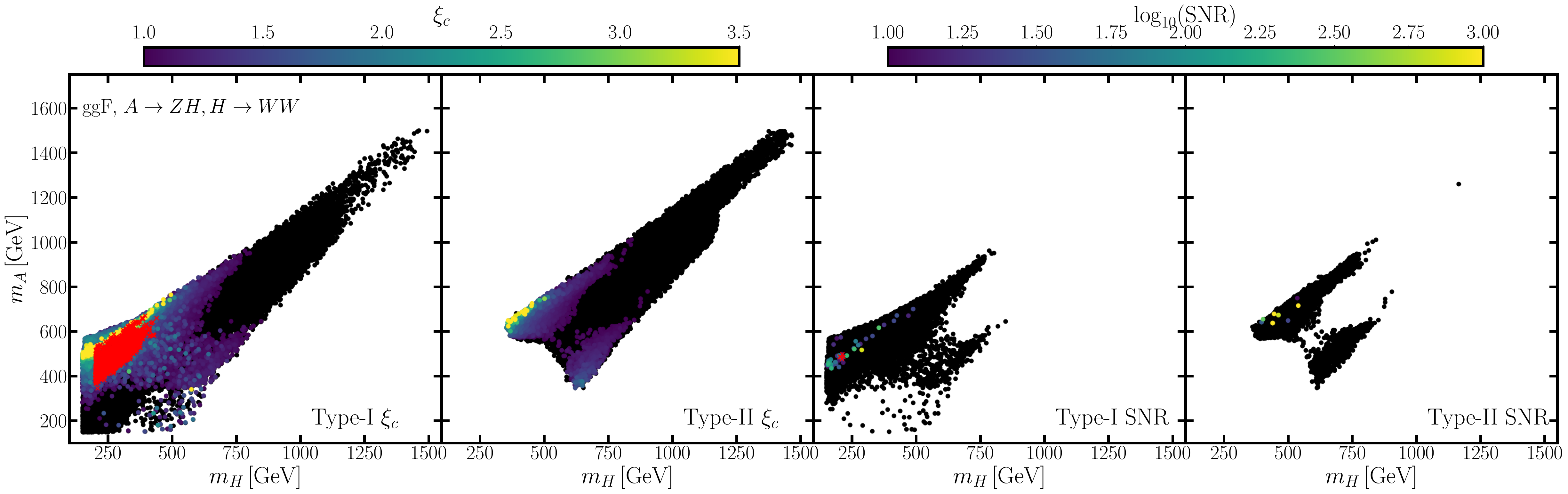}
    \caption{The $A\to ZH$ constraints on the $m_H$-$m_A$ plane. The red crosses are the points that can be probed by the HL-LHC through $A\to ZH\to \ell\ell b\bar b$ searches, where $A$ is produced through gluon fusion or via $b$-associated production (top). The projected HL-LHC $A\to ZH\to \ell\ell WW$  search is presented with red points, where $A$ is produced via gluon fusion (bottom).
	}
    \label{fig:AZH}
\end{figure}

Another important channel is $A\to ZH$, which is widely discussed in the context of EWPT in the 2HDM~\cite{Dorsch:2013wja,Dorsch:2014qja}. As we learned from~\autoref{sec:Vshape}, the left and top branch of~\autoref{fig:dma_dmh_xi_SNR} that corresponds to $m_A> m_H$ strongly favors the first-order phase transition due to the large and positive vacuum upliftment contributions from $A$ and possibly $H^\pm$. The favored parameter space dovetails nicely with the resonant searches through the $A\to ZH$ channel. Current experimental analyses explore this channel through the decays $H\to bb$ and $H\to WW$ with $Z\to\ell\ell$~\cite{Aad:2020ncx}. The corresponding constraints projected, according to the luminosity,  to the HL-LHC with $\mathcal{L}=3~\text{ab}^{-1}$ are shown in~\autoref{fig:AZH} for different channels. We observe that the $H\to b b$ channel (top panel) provides relevant limits, whereas the $H\to WW$ mode (bottom panel) results in smaller sensitivity, as it is suppressed by $c_{\beta-\alpha}$. In the type-II scenario, the latter channel does not provide extra sensitivity at the HL-LHC to SFOEWPT. This is due to the stronger constraints on $c_{\beta-\alpha}$ in the type-II 2HDM, pushing it further toward the alignment limit (see, e.g.~\autoref{fig:cba_MHH_xi_SNR} or~\autoref{fig:CHHH_cba_xi_SNR}). Remarkably, even exploring the dominant channel, where we have $A\to ZH$ and $H\to bb$, the sensitivity is still somewhat weak, being limited mostly to the parameter space region $m_H\lesssim 350$~GeV. The main reason is that the bottom quark decay channel quickly becomes subdominant once the scalar mass is beyond the top-pair threshold.\footnote{In the type-II scenario, the sensitivity to $H\to bb$ can extend beyond $350$~GeV, as large $t_\beta$ can enhance the branching ratio $\mathcal{BR}(H\to bb)$  even above the top-pair threshold.} Therefore, the fermionic channel with the top quark will be more promising in the high-mass region. We will explore the heavy scalar decays to top pairs in the next subsection.

The flipped channel $H\to ZA$ can \emph{a priori} also provide strong limits~\cite{Sirunyan:2019wrn}. It corresponds to the mass regime $m_H>m_A$, which is associated with the right and bottom branches of~\autoref{fig:dma_dmh_xi_SNR}. In the right branch, the SFOEWPT is suppressed due to limited positive contributions from heavy scalars to $\mathcal{F}_0$. Conversely, the bottom branch could still provide $\xi_c>1$ points with $t_\beta\to1$, as discussed in the previous section. However, this $t_\beta$ regime precludes  possible enhancements in the $A\to b\bar b$ branching fraction.
Thus, we do not observe extra sensitivity from this channel to the first-order phase transition at the HL-LHC.

\subsection{Scalar decays to heavy fermions}
\label{sec:Hff}

\begin{figure}[!tb]
    \centering
    \includegraphics[width=\textwidth]{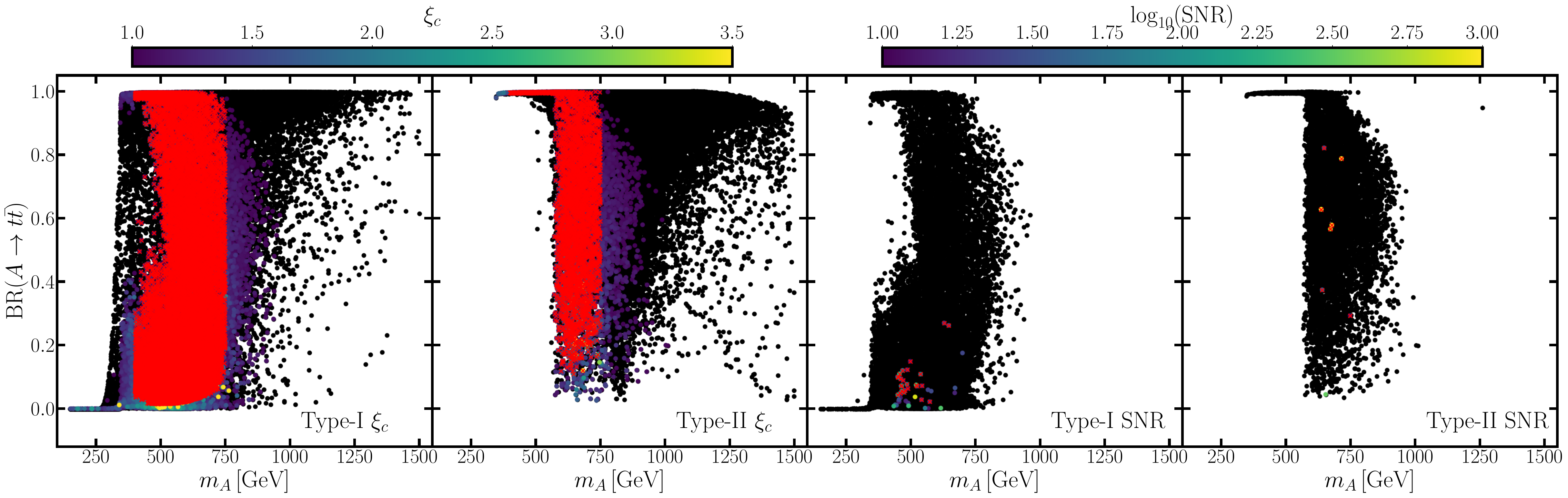}
    \includegraphics[width=\textwidth]{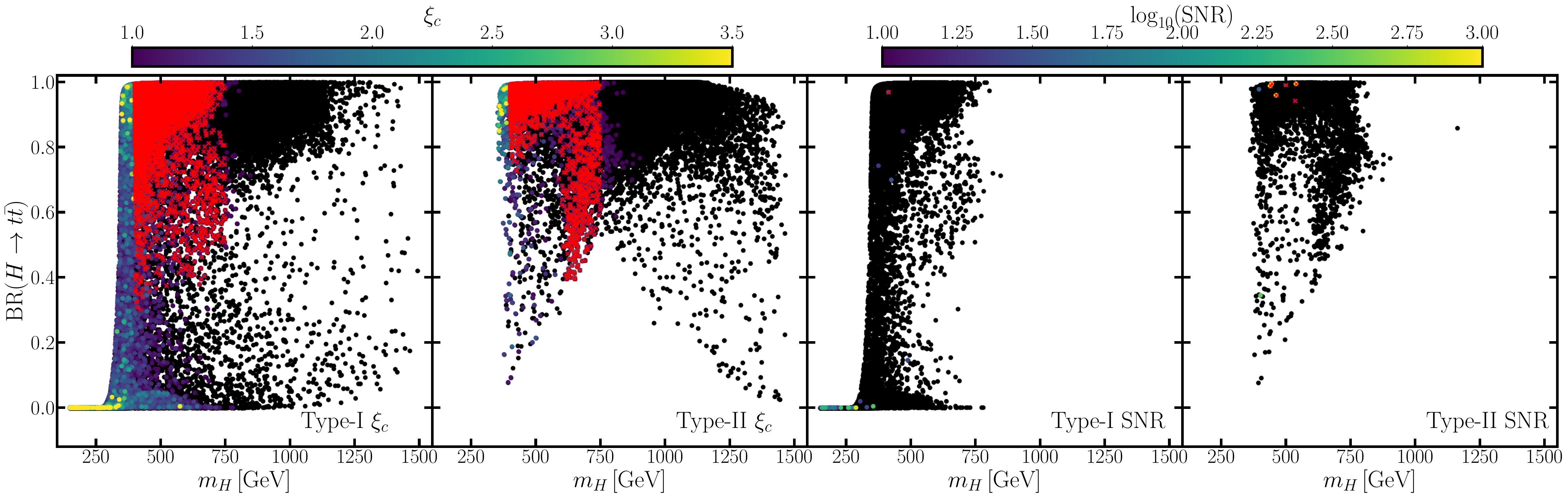}
    \caption{Branching fraction ${\mathcal{BR}}(A\to t\bar t)$ as a function of $m_A$ (top) and ${\mathcal{BR}}(H\to t\bar t)$ as a function of $m_H$ (bottom). The red crosses are the points with $\xi_c>1$ (left panels) and $\text{SNR}>10$ (right panels) that can be probed by the HL-LHC through resonant searches in the top quark pair final state. The color code is the same as in~\autoref{fig:dma_dmh_xi_SNR}.
    \label{fig:gAtt_MHA_xi_SNR}
    }
\end{figure}

\begin{figure}[!t]
    \centering
    \includegraphics[width=\textwidth]{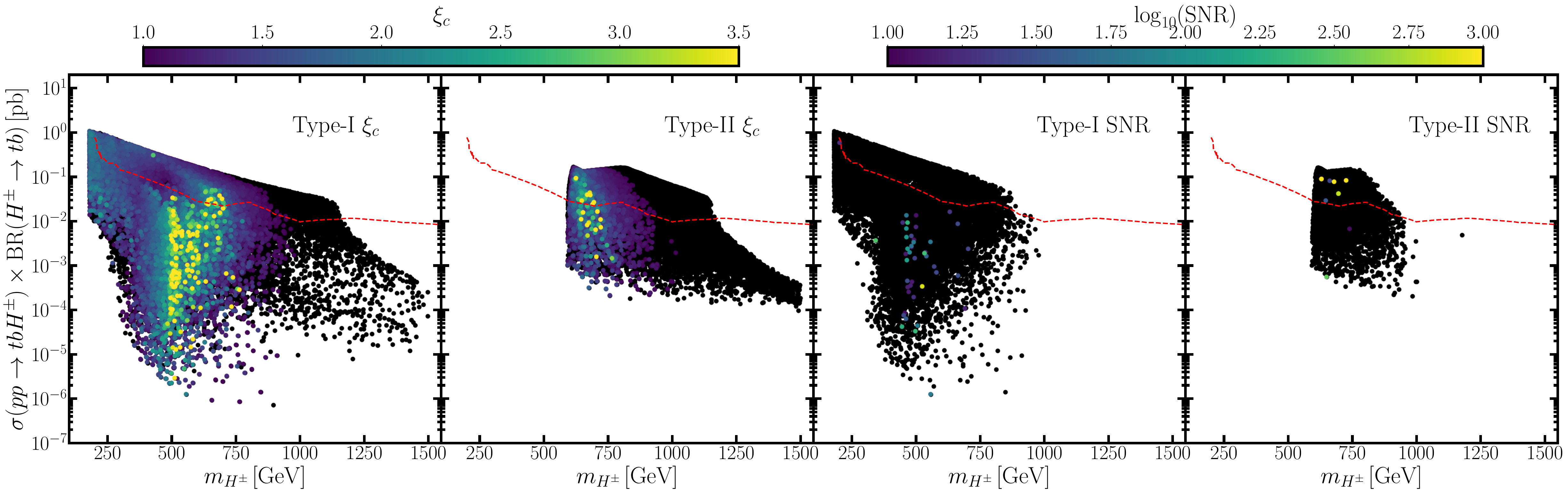}
    \caption{The cross section $\sigma(pp\to H^\pm tb)\times{\mathcal{BR}}(H^\pm \to tb)$ vs. $m_{H^{\pm}}$. The red dashed line indicates the projected limits from the HL-LHC. The color code is the same as~\autoref{fig:dma_dmh_xi_SNR}.}
    \label{fig:CS_Hpm_tb_MHp_xi_SNR}
\end{figure}

As observed above, resonant searches with heavy fermionic final states can be crucial for SFOEWPT sensitivity at the HL-LHC. Here, we discuss the projected constraints for both neutral scalars and charged Higgs decays. For the first, we consider the neutral scalar to top-pair search $H/A\to t\bar t$ performed by CMS~\cite{Sirunyan:2019wph} and scale it, according to the luminosity, to the HL-LHC.  Interestingly, this search can display large interference effects between the scalar mediated top-pair production and the SM background; thus, the sensitivity depends on the width of the relevant scalar~\cite{Dicus:1994bm}. The CMS experiment provided the likelihood as a function of the scalar coupling to the top pair for several choices of the scalar mass and width. We scale the likelihood according to the luminosity and linearly interpolate for the scalar mass and width to obtain the upper bound on the coupling for our parameter points.
In~\autoref{fig:gAtt_MHA_xi_SNR}, we show the branching fraction of $A$ and $H$ decaying into top quarks. The red crosses are the points that can be probed by the HL-LHC. From these plots, we find that the top-pair searches provide a promising search channel to probe SFOEWPT. The $H/A\to t\bar{t}$ searches will have special importance in the type-II 2HDM, as this scenario presents strong lower bounds on the scalar masses.

For the  charged scalars $H^\pm$, the main search channel at the LHC is  the charged scalar associated production with the top and bottom quarks $pp\to H^\pm tb$, where the $H^\pm$  subsequently decays into the top and bottom quarks $H^\pm \to tb$~\cite{Aad:2021xzu}. In~\autoref{fig:CS_Hpm_tb_MHp_xi_SNR}, we show the respective cross section times the branching ratio as a function of $m_{H^\pm}$. The red dashed line indicates the projected limits at the HL-LHC obtained by scaling the current bounds from Ref.~\cite{Aad:2021xzu} according to the luminosity to the HL-LHC with $3~\text{ab}^{-1}$. We observe that this channel is capable of covering the relevant region of the parameter space that can trigger strong first-order EWPT and  produce detectable gravitational wave signals.

\subsection{Combined results}

\begin{figure}[!tbp]
    \centering
    \includegraphics[width=\textwidth]{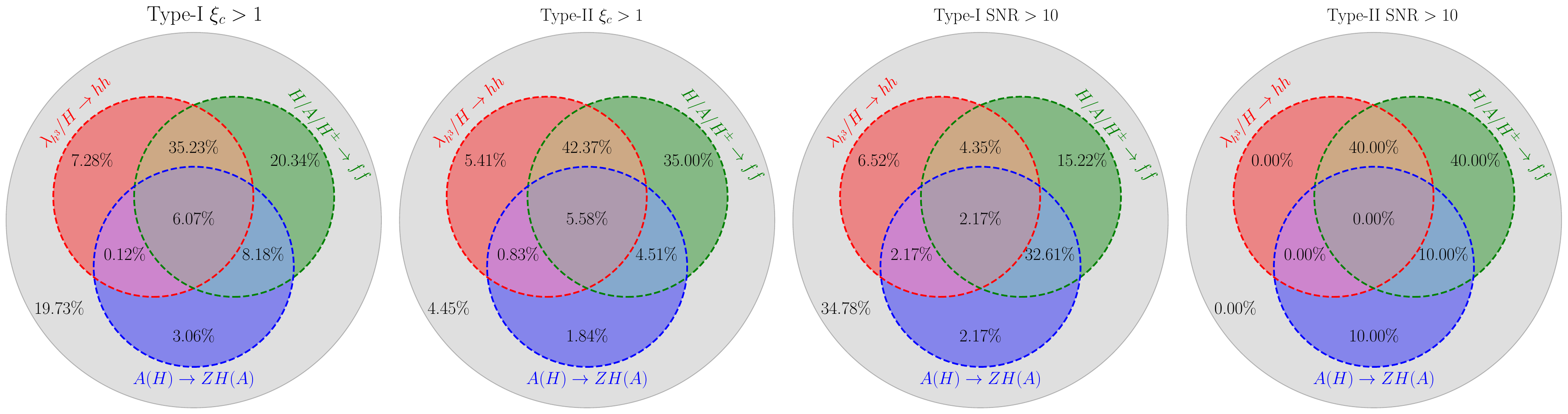}
    \caption{The summary of the capabilities of the corresponding search channels at the HL-LHC. The number in each region indicates the fraction of parameter points, currently allowed by theoretical and experimental constraints from our scan in that particular region.
    }
	\label{fig:Excl_Venn}
\end{figure}

\begin{figure}[!t]
	\centering
	\includegraphics[width=0.55\textwidth]{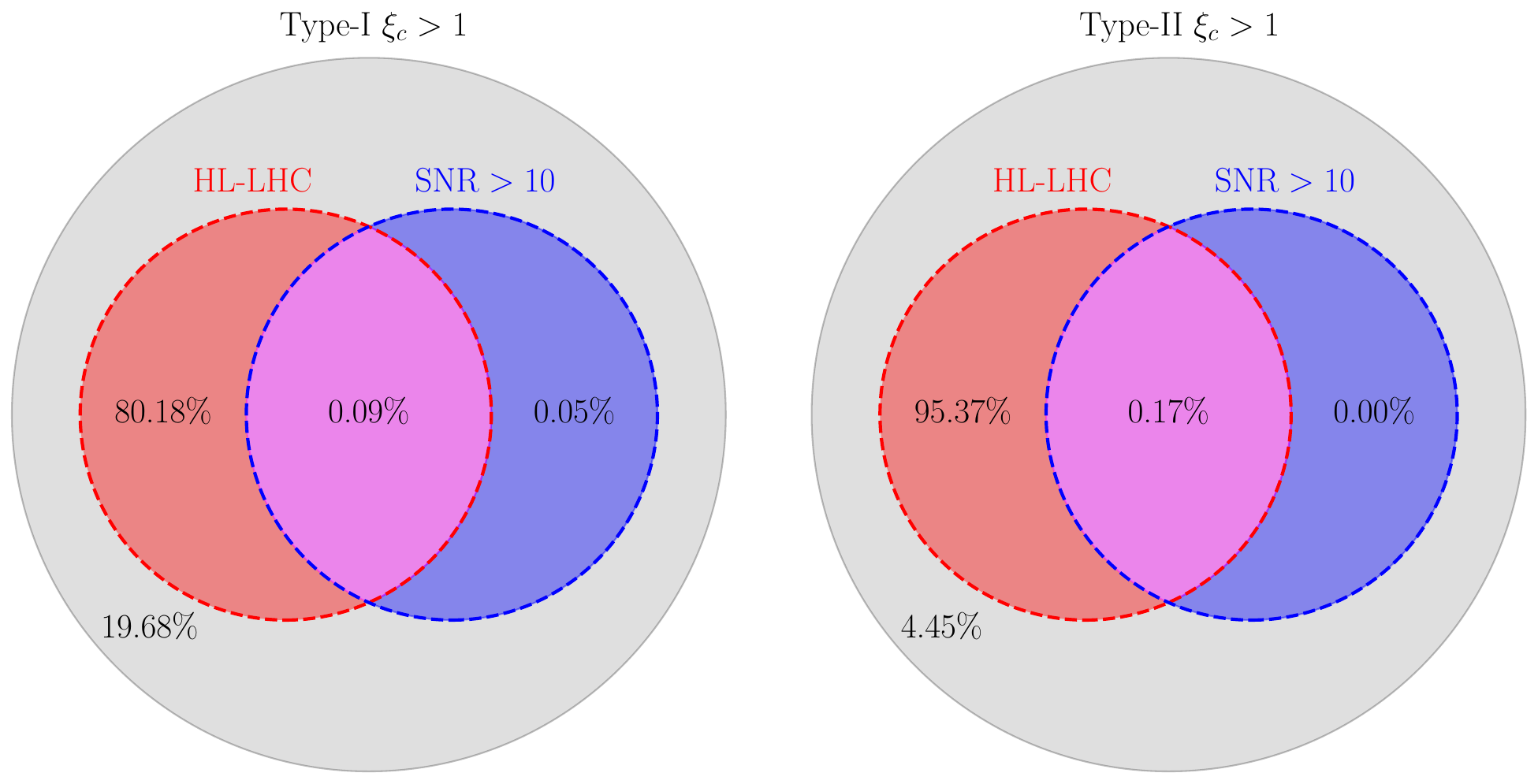}
	\caption{The summary of the capabilities of the HL-LHC and GW experiments. The number in each region indicates the fraction of parameter points from our scan in that particular region.}
	\label{fig:Complementarity_Venn}
\end{figure}

Here, we compare the sensitivity to SFOEWPT for the three aforementioned search channel categories:
\begin{itemize}
	\item Resonant and nonresonant di-Higgs
    \item $A\to ZH$ and $H\to ZA$
	\item $H\to t\bar t$, $A\to t\bar t$, and $H^\pm\to tb$
\end{itemize}
In~\autoref{fig:Excl_Venn}, we show the fractions of the parameter points from our uniformly random scan that can be covered by distinct search channels (or a combination of them). The left two panels show the case with $\xi_c>1$, while the right two panels show the case with ${\rm SNR}>10$. The percentage number in each subregion indicates the fraction of the parameter points in that specific subregion. From these Venn diagrams, one can clearly see that the above three categories of search channels are complementary to each other and  together can cover a wide portion of the remaining parameter space with strong first-order EWPT. Distinctly, we find that the fermionic modes as well as the di-Higgs channel provide the strongest sensitivity to SFOEWPT and complementary GW signals. For instance, when considering the possibility of probing the large order-parameter $\xi_c>1$, we observe that the combination of the di-Higgs mode with heavy fermionic decay channels can cover 77\% of the remaining type-I points and 94\% for the type-II points. At the same time, the widely discussed channel $A(H)\to ZH(A)$ is still relevant, but probes a smaller portion of the parameter space. The sensitivity from $A\to ZH$ could be, in principle, further enhanced by accounting for the $H\to t\bar{t}$ channel that is not yet performed by ATLAS and CMS. We leave this phenomenological study for future work~\cite{Goncalves:2022wbp}.

In~\autoref{fig:Complementarity_Venn}, we present the global HL-LHC and GW complementarities to probe the currently available SFOEWPT parameter space based on our uniformly random scan. We see that the HL-LHC searches will be able to cover $\approx 80\%$ of the remaining $\xi_c>1$ parameter space for the type-I 2HDM and an impressive $\approx 96\%$ for the type-II scenario.
In the type-I 2HDM, LISA provides additional sensitivity to a small complementary region where the production cross sections for the considered HL-LHC channels are suppressed. No such LISA only points are found in our type-II scan, although small complementary regions may remain unresolved because of the finite scan statistics. The requirement for small scalar masses to induce positive contributions for $\mathcal{F}_0$ plays a crucial role in the sizable HL-LHC sensitivity to SFOEWPT. These fractions represent only a lower bound. Adding other complementary 2HDM search channels at the HL-LHC, beyond the three considered classes, should push the quoted sensitivities to an even higher level.

\section{Summary}
\label{sec:summary}

Reconstructing the shape of the Higgs potential is crucial for understanding the origin of mass and the thermal history of electroweak symmetry breaking in our Universe. In this work, we explore the complementarity between collider and gravitational wave experiments to probe the scalar potential in the 2HDM. We scrutinize fundamental ingredients in the profile of the Higgs potential, namely the barrier formation and upliftment of the true vacuum that promote the transmutation of phase transition from the smooth crossover to the strong first-order phase transition. In addition, accounting for the theoretical and current experimental constraints, we study the prospects for the HL-LHC to probe the $\xi_c>1$ regime focusing on three prominent classes of searches (resonant and nonresonant di-Higgs, $A(H)\to ZH(A)$, and heavy scalar decays to fermions) and contrast them with the GW sensitivity at LISA. We summarize our novel results as follows:
\begin{itemize}
    \item When comparing the parameter space points that survive the theoretical and experimental constraints for type-I and type-II 2HDM, these scenarios result in an akin phase transition pattern.
    \item The barrier formation in the Higgs potential of the 2HDM is driven by the one-loop and thermal corrections, with the dominance of the one-loop terms for large order-parameter $\xi_c>1$.
    \item The strength of phase transition is correlated with the upliftment of the true vacuum with respect to the symmetric one at zero temperature~\cite{EWPT-NMSSM, EWPT-Nature,Dorsch:2017nza}. This arises as a result of the dominance of the one-loop effects with respect to the thermal corrections for $\xi_c>1$. Based on this result, we shed light on the phase transition pattern analytically. In particular, we observe that larger vacuum upliftment is favored for lower scalar masses {which is in accordance with the results from a generic discussion in~\cite{Ramsey-Musolf:2019lsf} }. This provides strong extra motivation for scalar searches at the LHC. Besides scalar masses below the TeV scale, the analytical structure of the new physics effects on the vacuum upliftment, leading to SFOEWPT, results in a peculiar hierarchy of masses among the new scalar modes.
    These findings work as a guide for collider and gravitational wave studies.
    \item We obtain that the scalar decays to heavy fermions $(H,A,H^\pm\to tt, tb)$ are the most promising smoking gun signature for SFOEWPT at the HL-LHC, followed by the di-Higgs searches. Based on the projections from the current ATLAS and CMS searches, the widely discussed channel $A(H)\to ZH(A)$ is still relevant, whereas to a smaller fraction of the parameter space. The main reason for such an observation is that the current experiments focus on the $bb$ and $WW$ decay channels~\cite{Aad:2020ncx}. These two decay modes only cover a small portion of the parameter space. We leave for future work a direct phenomenological comparison of the gluon fusion  $gg\to H(A)$ and $A(H)\to ZH(A)$ channels, considering the promising $t\bar t$ heavy fermion final states~\cite{Goncalves:2022wbp}.
    \item In contrast to the HL-LHC, LISA is sensitive to a significantly smaller region of parameter space. In the Type-I 2HDM it provides complementary sensitivity in regions where the corresponding LHC cross sections are suppressed. No such complementary LISA sensitivity is found in the Type-II 2HDM, for which the HL-LHC is generally more restrictive. Based on our parameter-space scan, the combination of LHC searches and gravitational wave observations offers promising prospects for probing the vast majority of first-order phase transition points in the 2HDM. Including additional complementary 2HDM search channels at the HL-LHC, beyond the three classes considered here,  should push new physics sensitivity to an even higher level~\cite{Goncalves:2022wbp}.
\end{itemize}

 In conclusion, the study of the thermal history of electroweak symmetry breaking is a crucial challenge for particle physics and cosmology. We demonstrate that the well-motivated 2HDM leads to a rich phase transition pattern favoring SFOEWPT below the TeV scale. This renders exciting physics prospects at the HL-LHC and upcoming gravitational wave experiments, such as LISA.

\section*{Acknowledgements}
\label{sec:acknowledgements}
We thank Maria Cepeda and Alessia Saggio for clarifying some details of Ref.~\cite{Sirunyan:2019wrn}. D.G., A.K., and Y.W. thank the U.S.~Department of Energy for the financial support, under Grant No. DE-SC 0016013. Some computing for this project was performed at the High Performance Computing Center at Oklahoma State University, supported in part by the National Science Foundation Grant No. OAC-1531128.

\appendix
\section{Potential Parameters in the 2HDM}
\label{app:parameters}
In this appendix, we express the masses $m_{11}^2$, $m_{22}^2$ and coupling parameters $\lambda_1....\lambda_5$ in terms of the parameters  $m_h$, $m_H$, $m_A$, $m_{H^\pm}$, $\beta$, $\alpha$ and $m_{12}^2$ used throughout this manuscript:
\begin{subequations}
 \begin{align}
     \lambda_1v^2 &= \frac{1}{c_\beta^2}\left(s_\alpha^2 m_h^2 + c_\alpha^2 m_H^2 - m_{12}^2\tan\beta\right), \\
     \lambda_2v^2 &= \frac{1}{s_\beta^2}\left(c_\alpha^2 m_h^2 + s_\alpha^2 m_H^2 - m_{12}^2/\tan\beta\right),\\
     \lambda_3v^2 &= 2m_{H^\pm}^2+\frac{s_{2\alpha}}{s_{2\beta}}(m_H^2-m_h^2)-\frac{m_{12}^2}{s_\beta c_\beta},\\
     \lambda_4v^2 &= m_A^2 - 2m_{H^\pm}^2+\frac{m_{12}^2}{s_\beta c_\beta},\\
     \lambda_5v^2 &= \frac{m_{12}^2}{s_\beta c_\beta}-m_A^2,\\
     m_{11}^2 &= m_{12}^2\tan\beta - \frac{m_H^2}{2}\frac{c_\alpha}{c_\beta}c_{\beta-\alpha} + \frac{m_h^2}{2}\frac{s_\alpha}{c_\beta}s_{\beta-\alpha},\\
     m_{22}^2 &= \frac{m_{12}^2}{\tan\beta} - \frac{m_H^2}{2}\frac{s_\alpha}{s_\beta}c_{\beta-\alpha} - \frac{m_h^2}{2}\frac{c_\alpha}{s_\beta}s_{\beta-\alpha}.
 \end{align}
\end{subequations}

\bibliographystyle{JHEP}
\bibliography{references}
\end{document}